\documentclass[%
reprint, superscriptaddress, amsmath, amssymb, aps, prx]{revtex4-2}
\usepackage{amsmath}
\usepackage[yyyymmdd]{datetime}
\usepackage{bm}
\usepackage{color}
\usepackage{colortbl}
\usepackage{dcolumn}
\usepackage{graphicx}
\usepackage{physics}
\usepackage{hyperref}
\usepackage{sidecap}
\usepackage[dvipsnames]{xcolor}
\usepackage{xspace}
\usepackage{siunitx}
\usepackage{float}

\DeclareSIUnit\gauss{G}
\hypersetup{ 
    colorlinks=true,
    linkcolor=blue,     
    urlcolor=blue,
    citecolor=blue,
}

\begin{document}
\preprint{APS/123-QED}
\title{Continuously trapped matter-wave interferometry\\ in magic Floquet-Bloch band structures}
\author{Xiao Chai}
\author{Eber Nolasco-Martinez}
\thanks{Equal contribution.}
\author{Xuanwei Liang}
\thanks{Equal contribution.}
\author{Jeremy L. Tanlimco}
\thanks{Equal contribution.}
\author{E. Quinn Simmons}
\author{Eric Zhu}
\author{Roshan Sajjad}
\author{Hector Mas}
\author{S. Nicole Halawani}
\author{Alec Cao}
\author{David M. Weld}
\email{weld@physics.ucsb.edu}
\affiliation{Department of Physics, University of California, Santa Barbara, California 93106, USA}

\begin{abstract}
Trapped matter-wave interferometry offers the promise of compact high-precision local force sensing. However, noise in the trap itself can introduce new systematic errors which are absent in traditional free-fall interferometers. We describe and demonstrate an intrinsically noise-tolerant Floquet-engineered platform for continuously trapped atom interferometry. A non-interacting degenerate quantum gas undergoes position-space Bloch oscillations through an amplitude-modulated optical lattice, whose resulting Floquet-Bloch band structure includes Landau-Zener beamsplitters and Bragg mirrors, forming the components of a Mach-Zehnder interferometric force sensor. We identify, realize, and experimentally characterize magic band structures, analogous to the magic wavelengths employed in optical lattice clocks, for which the interferometric phase is insensitive to lattice intensity noise. We leverage the intrinsic programmability of the Floquet band synthesis approach to demonstrate a variety of interferometer structures, highlighting the potential of this technique for quantum force sensors which are tunable, compact, simple, and robust.
\end{abstract} 


\maketitle
\let\oldaddcontentsline\addcontentsline
\renewcommand{\addcontentsline}[3]{}

\section{Introduction}
Interferometry has been a tool of discovery in physics for centuries~\cite{young_i_1804,abbott_observation_2016}. While originally demonstrated with photons, the wave-particle duality~\cite{de_broglie_recherches_1925} of matter enables the realization of interferometers in electrons~\cite{marton_electron_1953}, neutrons~\cite{rauch_test_1974}, atoms~\cite{baudon_atomic_1999}, and molecules~\cite{borde_molecular_1994}. The interferometric sensitivity is set by the spacetime area enclosed by the interferometer loop and thus scales quadratically with freefall time for untrapped particles. This has driven the precision frontier towards large-scale experiments featuring 100-meter-scale drop towers \cite{abe_matter-wave_2021} or even low Earth orbit \cite{elliott_quantum_2023}. In contrast, continuously-trapped interferometers can reach very large spacetime areas in the Earth's gravity field without requiring long freefall time, large experimental size, or spaceflight, and offer the practical advantages of compactness and truly local sensing~\cite{schumm_matter-wave_2005,wu_demonstration_2007,garcia_bose-einstein-condensate_2006,mcdonald_optically_2013,panda_measuring_2024, weidner_experimental_2018,ledesma_vector_2025,petrucciani_mach-zehnder_2025,li_matter-wave_2024}. 
However, instabilities in the trapping potential result in dephasing. While this can be partly mitigated by resonant optical mode filtering~\cite{xu_probing_2019} or differential measurement schemes~\cite{petrucciani_mach-zehnder_2025}, such dephasing still represents a primary limitation of trapped matter-wave interferometers.

\begin{figure*}
    \centering
    \includegraphics[width=18cm]{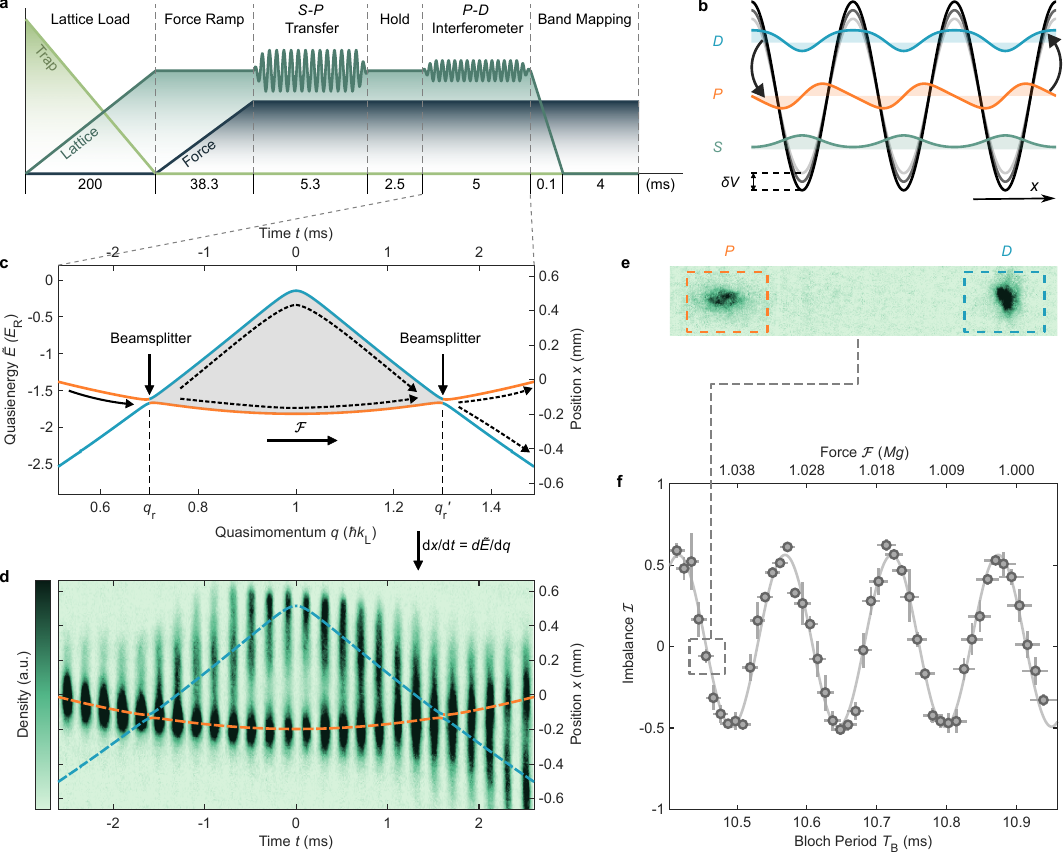}
    \caption{\textbf{Floquet-Bloch matter-wave interferometry.} (\textbf{a}) Experimental sequence. (\textbf{b}) Resonant amplitude modulation of the optical lattice hybridizes the $P$ and $D$ (second and third) Bloch bands. (\textbf{c}) Interferometric loop formed by the Floquet-Bloch band structure and Landau-Zener beamsplitters. A matter wave Bloch oscillates through the paths indicated by the arrows in response to a force $\mathcal{F}$. The color of the Floquet-Bloch bands corresponds to the Bloch states in (b) with  maximal wavefunction overlap. (\textbf{d}) Time sequence of \textit{in} \textit{situ} images of a condensate traversing the interferometric loop. The space-time trajectories map out the Floquet-Bloch band structure. (\textbf{e}) Band-mapping spatially separates the two output port populations. (\textbf{f}) Output population imbalance $\mathcal{I}$ versus applied force. $Mg$ is the mass times the local gravitational acceleration, used here only as a force scale; the lattice is horizontal. In all figures, vertical error bars are the standard error of 3 repeated measurements and horizontal error bars are the estimated uncertainties from calibration fits \cite{sup}.}
    \label{fig:1}
\end{figure*}

In this work we describe and experimentally characterize a new and highly flexible class of atom interferometer based on position-space Bloch oscillations~\cite{dahan_bloch_1996,geiger_observation_2018} of trapped non-interacting Bose-condensed lithium through loops in a Floquet-engineered optical lattice band structure \cite{holthaus_floquet_2016,fujiwara_transport_2019,weitenberg_tailoring_2021,sandholzer_floquet_2022}. The sensitivity of this interferometer scales inversely with the applied force, making it especially suitable for detection of weak forces. Instead of using traditional Raman or Bragg pulses, matter-wave splitting and recombination is realized via tunable Landau-Zener transitions at quasimomentum-selective Floquet-Bloch band crossings. This makes the interferometer intrinsically insensitive to fluctuations in initial momentum, pulse duration, and laser phase. Immunity against trap intensity fluctuations is achieved by the use of an infinite family of magic Floquet-Bloch band structures, which generalize the concepts of magic wavelengths and magic lattice depths~\cite{mcalpine_excited-band_2020}. We experimentally verify that magic Floquet-Bloch bands exhibit first-order insensitivity to lattice amplitude noise, without requiring mode filtering with a resonant cavity. These advantages are a consequence of the remarkable degree of design flexibility offered by Floquet band engineering, which takes inspiration from related concepts in condensed matter physics~\cite{basov_towards_2017,oka_floquet_2019}
and photonics~\cite{wang_photonic_2023,zhao_blochzener_2024} and enables the synthesis of a wide variety of interferometer loops with different characteristics and tunable force response.

\section{Working Principle}
The central component of the interferometers we describe is a horizontal one-dimensional optical lattice amplitude-modulated to hybridize the Bloch bands (Fig.~\ref{fig:1}). The  lattice potential takes the form
$V(x,t) =  - \left(V_0+\delta V \sin\omega t \right) \cos^2(k_\mathrm{L} {x})$, 
where $V_0$ is the static lattice depth, $\delta V$ and $\omega$ are the modulation depth and frequency respectively, $k_\mathrm{L} = 2\pi/\lambda$ is the lattice laser wavenumber, and $\lambda = \SI{1064}{\nano\meter}$ is the lattice laser wavelength. This yields a spatially and temporally periodic single-particle Hamiltonian to which Floquet-Bloch theory can be readily applied \cite{holthaus_floquet_2016}. As shown in Fig.~\ref{fig:1}c, the modulation opens up gaps between Floquet-Bloch quasienergy bands when the resonance condition $E_{n,q_{\mathrm{r}}}-E_{n',q_{\mathrm{r}}} = \hbar \omega$ is satisfied, where $E_{n,q}$, $E_{n',q}$ are the bare Bloch band energies for bands labeled $n$ and $n'$ (henceforth $S,P,D$, etc., following orbital notation), $q$ is the quasimomentum, $q_\mathrm{r}$ is the resonant quasimomentum, and $\hbar$ is the reduced Planck constant. When a matter wave in one band traverses the avoided crossing at $q=q_\mathrm{r}$, it coherently splits into a superposition state between the two output bands~\cite{landau1932theorie1,landau1932theorie2,zener_non-adiabatic_1932}.
By tuning the modulation depth $\delta V$, the Landau-Zener transition probability $p$ can be adjusted to 50\%~\cite{cao_probing_2020}.
Arbitrary synthesis of the RF modulation waveform allows fully programmable manipulation of both the location and strength~\cite{cao_probing_2020} of all such avoided crossings between quasienergy bands: these will become the beamsplitters and other control elements comprising the interferometer.   

In the presence of a force $\mathcal{F}$ along the lattice direction, the acceleration theorem \cite{arlinghaus_generalized_2011} guarantees that the quasimomentum $q$ of a matter wave evolves as $q(t) = q(0) + \mathcal{F}t$, with the Bloch period defined by traversal of the  entire Brillouin zone:
\begin{align}
    T_\mathrm{B} = \frac{2\hbar k_\mathrm{L}}{\mathcal{F}}.
    \label{eqn:Bloch_period}
\end{align}
Crucially, this St\"uckelberg-type evolution in quasienergy and quasimomentum~\cite{kling_atomic_2010,zenesini_observation_2010}  also results in motion in real space: the mean position $x(t)$ of a wavepacket in a Floquet-Bloch band with quasienergy dispersion relation $\tilde{E}(q)$ evolves according to the group velocity as $\dd{x}/\dd{t} = \dd{\tilde{E}}/\dd{q}$~\cite{geiger_observation_2018}. The displacement associated with such position-space Bloch trajectories can be especially large for light atoms like lithium; separations on the millimeter scale are straightforwardly attained. This means that by adjusting the modulation waveform to synthesize a particular Floquet-Bloch band structure between two Landau-Zener avoided crossings, we are synthesizing a particular spacetime trajectory for the atoms in the interferometer loop: the quasienergy band dispersion is simply a scaled image of the center of mass time evolution~\cite{fujiwara_transport_2019}, as shown in Fig.~\ref{fig:1}c-d. Components of the atomic wavefunction which are separated at an initial Landau-Zener beamsplitter and traverse the two arms of such a loop accumulate a relative dynamical phase which depends on the applied force. This phase is measured by interfering the two pathways at the second beamsplitter, forming a Landau–Zener–St\"uckelberg–Majorana interferometer \cite{shevchenko_landauzenerstuckelberg_2010,ivakhnenko_nonadiabatic_2023} and resulting in a final population imbalance between the two output bands \cite{sup}
$
    \mathcal{I} = p_\mathrm{L} - p_\mathrm{U} \approx C\cos\phi_\mathrm{Int},
$
where $\rm L(U)$ denotes the lower (upper) Floquet-Bloch band, $p_\mathrm{L(U)}$ the corresponding population fractions, and the ideal contrast $C=4p(1-p)$. The interferometer phase is
$
    \phi_\mathrm{Int} = \phi_\mathrm{Dyn}+2\phi_\mathrm{Sto},
$
where $\phi_\mathrm{Sto} \in [-\pi/2,-\pi/4]$ is the Stokes phase imparted by one of the Landau-Zener transitions, which depends only weakly on the applied force \cite{sup}. $\phi_\mathrm{Dyn}$ is the differential dynamical phase
\begin{align}
    \phi_\mathrm{Dyn} = \frac{T_\mathrm{B}}{2 \hbar^2 k_\mathrm{L}}\int_{q_\mathrm{r}}^{q'_\mathrm{r} = q_\mathrm{r}+\Delta q} \dd{q} \left( \tilde{E}_{\mathrm{U},q} - \tilde{E}_{\mathrm{L},q}\right),
    \label{eqn:dynamicalPhaseDifference}
\end{align}
where $\tilde{E}_{\mathrm{L}(\mathrm{U}),q}$ denotes the quasienergy. The dynamical phase is proportional to $T_\mathrm{B}$ and thus $1/\mathcal{F}$; the force sensitivity scales linearly with the energy-momentum area enclosed by the interferometer loop.

\begin{figure*}
    \centering
    \includegraphics[width=18cm]{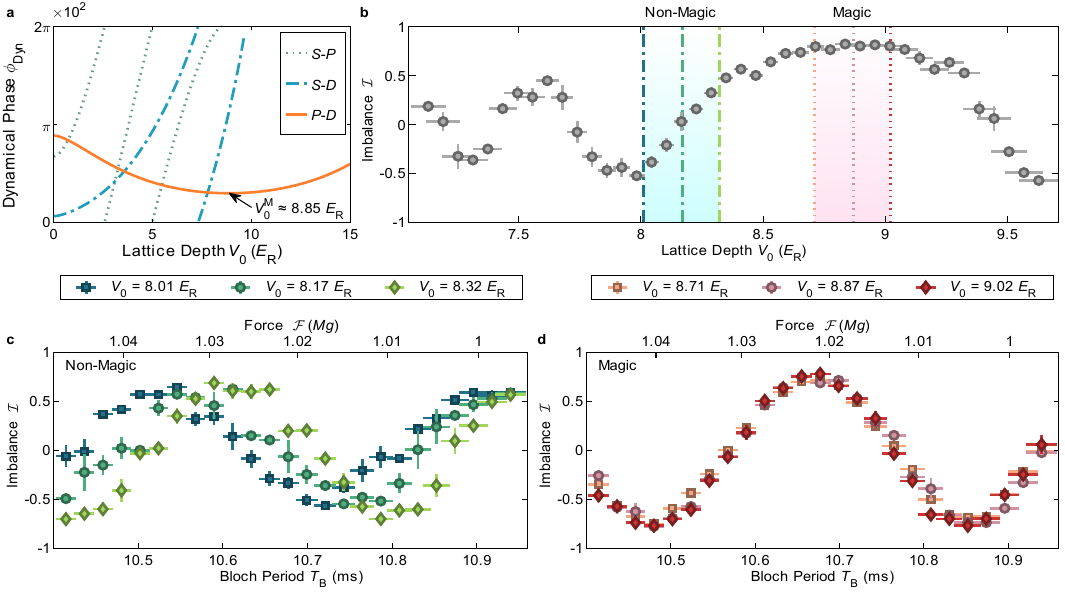}
    \caption{\textbf{Magic band structures.} (\textbf{a}) Numerically calculated dynamical phase (Eq.~\ref{eqn:dynamicalPhaseDifference}) modulo $2\pi\times10^2$ for $S$-$P$, $S$-$D$, and $P$-$D$ interferometric loops (with $\SI{100}{\kilo\hertz}$, $\SI{190}{\kilo\hertz}$, and $\SI{143}{\kilo\hertz}$ drives, respectively) as a function of lattice depth for $T_\mathrm{B}=10.7\,\mathrm{ms}$ estimated with static band energies; only the $P$-$D$ loop exhibits a magic condition at nonzero $V_0^\mathrm{M}\approx8.85\,E_\mathrm{R}$. (\textbf{b}) Output population imbalance at $T_\mathrm{B}=10.70\,\mathrm{ms}$ as a function of lattice depth. Vertical dashed (dotted) lines indicate the lattice depths chosen for force measurements far from (close to) the magic condition. (\textbf{c}-\textbf{d}) Imbalance as a function of applied force away from (c) and close to (d) the magic condition at the aforementioned lattice depths.}
    \label{fig:2}
\end{figure*}

\section{Experimental Implementation}
The experimental apparatus with which we realize this interferometer builds upon previous work (see \cite{fujiwara_transport_2019} and Appendix~\ref{app:exp}). A ${}^7 \rm{Li}$ Bose-Einstein condensate (BEC) is prepared in the $|f, m_f\rangle = |1,1\rangle$ magnetically sensitive state within a crossed dipole trap. To eliminate interaction-induced dephasing~\cite{petrucciani_mach-zehnder_2025}, we apply a magnetic field of $\SI{543.6}{\gauss}$ to tune the $s$-wave scattering length to zero \cite{pollack_extreme_2009} using lithium's shallow Feshbach zero-crossing. The atoms are then adiabatically loaded into the ground band of a horizontal one-dimensional optical lattice (Fig.~\ref{fig:1}a) with a lattice depth $V_0 = 8.45\,E_\mathrm{R}$, where $E_\mathrm{R} = \hbar^2 k_\mathrm{L}^2 / 2M$ is the recoil energy, and $M$ is the atomic mass. Subsequently, we ramp up a magnetic field gradient effecting a nearly uniform force $\mathcal{F}$ along the lattice direction to initiate Bloch oscillations. A strong $\SI{120}{\kilo\hertz}$ lattice amplitude modulation pulse is then applied to transfer the atoms from the $S$ to $P$ band with ~100\% Landau-Zener probability (see Appendix~\ref{app:sp}). This initial transfer allows us to implement the interferometer using $P$-$D$ hybridized Floquet-Bloch bands, the lowest combination which supports magic band structures. After a $\SI{2.5}{\milli\second}$ hold time, a $\SI{127.438}{\kilo\hertz}$ modulation is ramped up within $\SI{100}{\micro\second}$ to load the atoms adiabatically into the upper $P$-$D$ hybridized Floquet-Bloch band. The modulation depth is calibrated to be $\delta V = 0.35\,E_\mathrm{R}$ for 50-50 beam splitting \cite{sup}. The modulation is sustained for $\SI{4.8}{\milli\second}$, during which time the atoms traverse the Brillouin zone edge and complete the interferometer loop at the second crossing. The second crossing is created by the same modulation and therefore is mirrored around the zone edge from the first crossing; this allows the interferometer loop to close. Figure \ref{fig:1}d shows a series of snapshots capturing the atoms traversing the interferometer loop, where the center-of-mass motion clearly maps out the Floquet-Bloch band structure \cite{geiger_observation_2018,fujiwara_transport_2019}. Finally, the modulation is ramped down in $\SI{100}{\micro\second}$ to convert atoms in upper (lower) Floquet-Bloch bands back into the static $P$ ($D$) Bloch bands. We perform band-mapping \cite{greiner_exploring_2001} and absorption imaging to read out the band populations with high fidelity (see Fig.~\ref{fig:1}e and Appendix~\ref{app:data}). Figure~\ref{fig:1}f demonstrates a functioning interferometer, displaying a measured interference fringe in which the population imbalance depends sinusoidally on the inverse of the applied force $\mathcal{F}$.

\section{Magic Band Structure}
In addition to force, the differential dynamical phase in general also depends sensitively on lattice depth. If this were always the case it would spoil the performance of force sensors of this type~\cite{hilico_contrast_2015}. Fortunately, the flexibility and large available parameter space of multi-frequency Floquet band engineering allows for the design and creation of magic band structures which null out the lattice depth dependence. The condition for magicness of a Floquet-Bloch band loop is $\pdv*{\phi_\mathrm{Int}}{V_0}=0$. This condition cannot be realized in any simple loop whose constituent bands include the ground ($S$) band. The reason for this is that the perturbative effect of nonzero $V_0$ on the free-particle parabolic dispersion serves to repel the different bands; since the $S$ band uniquely only experiences repulsion from above, its energy monotonically decreases with increasing lattice depth (and more steeply than any other band), precluding the existence of any nonzero $V_0$ for which the energy difference with another band has zero derivative~\cite{mcalpine_excited-band_2020}. Thus, the simplest pair of bands which can realize a ``magic-depth'' interferometer are the $P$ and $D$ bands. For example, as demonstrated numerically in Fig.~\ref{fig:2}a, the magic condition is realized for a $P$-$D$ interferometric loop extending from $q_\mathrm{r}=0.8\,\hbar k_\mathrm{L}\to q_\mathrm{r}'=1.2\,\hbar k_\mathrm{L}$ at lattice depth $V_0^\mathrm{M}\approx8.85\,E_\mathrm{R}$. This approach can be generalized to different loop areas and interferometer geometries: such magic band structures can be found for essentially any quasimomentum range \cite{sup} and any set of constituent excited bands. 

\begin{figure*}
    \centering
    \includegraphics[width=18cm]{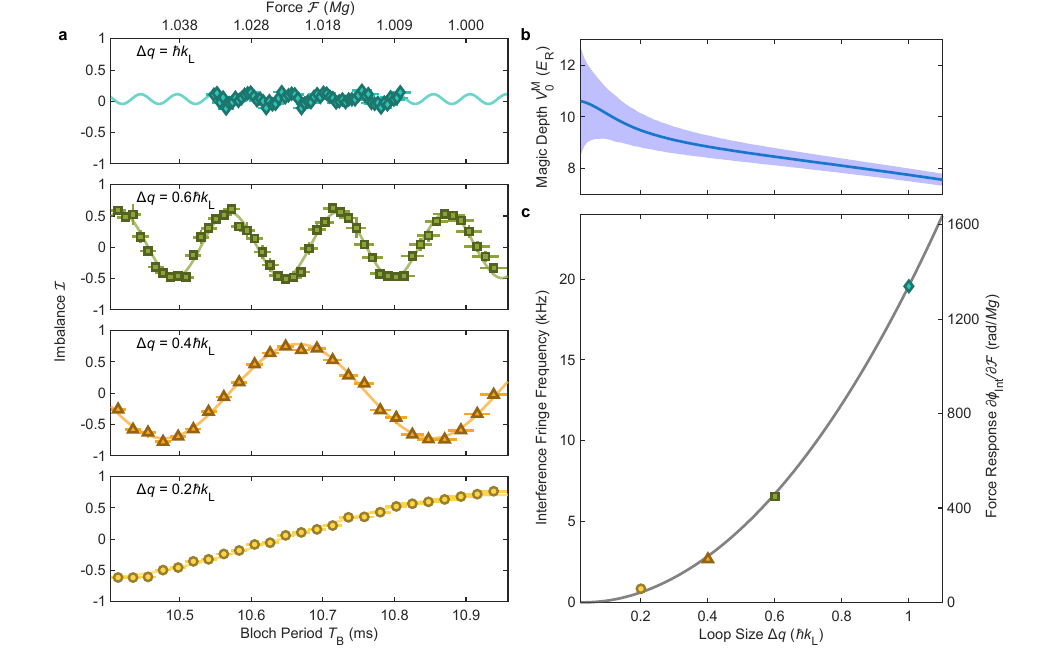}
    \caption{\textbf{Tuning the interferometer response}. (\textbf{a}) Interference fringes for varying loop sizes $\Delta q = q'_\mathrm{r} - q_\mathrm{r}$, each at the respective magic condition. Solid lines are sinusoidal fits. (\textbf{b}) Theoretically calculated magic depth as a function of loop size. Boundaries of the shaded area mark the lattice depth where the interferometer phase deviates by $\pi/4$. (\textbf{c}) Interferometric response as a function of loop size as defined by the fringe frequency. Data points represent fit results from (a), and the solid line is the theoretical prediction obtained from the fit-parameter-free analytical theory. The corresponding force response is calculated at $\mathcal{F} = Mg$.}
    \label{fig:3}
\end{figure*}

\begin{figure*}
    \centering
    \includegraphics[width=\linewidth]{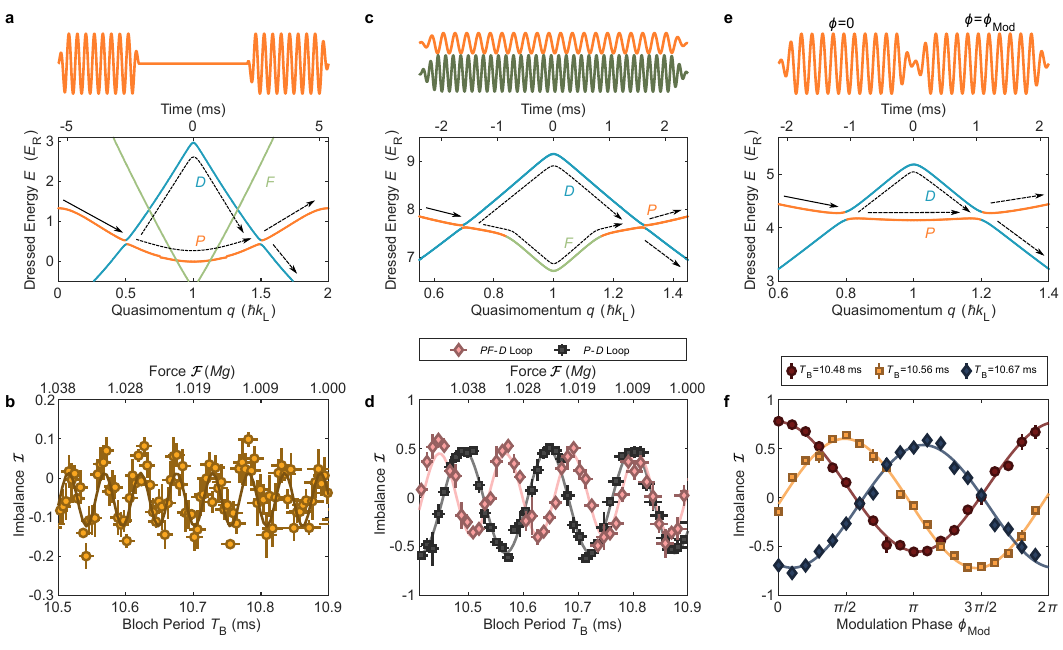}
    \caption{\textbf{Programmable Floquet synthesis of various interferometer structures.
    } 
    (\textbf{a}) Dressed energies of the $\Delta q = \hbar k_\mathrm{L}$ loop and the corresponding modulation waveform. The modulation is pulsed to avoid coupling with the \textit{F} band. (\textbf{b}) Interferometric fringe generated by the temporally separated pulses. (\textbf{c}) Dressed energies and modulation waveform of the $\Delta q=0.6\,\hbar k_\mathrm{L}$ loop in the presence of a second $196\,\mathrm{kHz}$ drive (dark green), which hybridizes the \textit{P} and \textit{F} bands. (\textbf{d}) Interferometric fringes with and without the second drive. (\textbf{e}) Dressed energies and modulation waveform of the $\Delta q=0.4\,\hbar k_\mathrm{L}$ interferometer loop. The phase of the second pulse is shifted by $\phi_\mathrm{Mod}$ relative to the first. (\textbf{f}) Output population imbalance as a function of $\phi_\mathrm{Mod}$ for three different forces.
    }
    \label{fig:4}
\end{figure*}

To test the predicted magic band structure, we first measure the population imbalance at fixed force and modulation waveform across a range of lattice depths near this value. The results, shown in Fig.~\ref{fig:2}b, indeed indicate a broad region with $\pdv*{\phi_\mathrm{Int}}{V_0}\approx0$ centered on $V_0=8.85\,E_\mathrm{R}$. To probe directly the performance of magic versus non-magic band structures, we measure force scans for three closely spaced lattice depths away from the magic condition (Fig.~\ref{fig:2}c) and for three closely spaced lattice depths centered at the magic condition (Fig.~\ref{fig:2}d).  Measurements far from the magic condition exhibit sensitive dependence on lattice depth and greater fluctuations in imbalance for a given lattice depth; in contrast, force measurements at or near the magic condition are quiet and consistent across a range of lattice depths. These results clearly demonstrate both the successful experimental implementation of magic Floquet-Bloch band synthesis and its utility for force sensing.

\section{Force Response}
One key motivation for continuously trapped atom interferometry is the ability, in principle unlimited by geometrical constraints, to scale up the spacetime area of the interferometer loop to increase sensitivity. To probe such scaling, we use the intrinsic flexibility of Floquet-synthesized interferometry to create a sequence of interferometers with increasing spacetime loop area. Specifically, we measure interference fringes produced using loops which subtend spans of quasimomenta $\Delta q=q'_\mathrm{r} - q_\mathrm{r}$ ranging from $0.2\,\hbar k_\mathrm{L}$ to $1\,\hbar k_\mathrm{L}$, keeping the interferometer loop symmetric about the Brillouin zone edge. This probes the performance of interferometers with different momentum-space beamsplitter separations. For each loop size, we recalculate the magic condition and adjust the lattice depth accordingly. At the magic condition, the range of acceptable lattice depth fluctuations decreases with increasing loop size (Fig.~{\ref{fig:3}}b), but for all measurements remains comfortably above the experimentally achievable lattice depth stability. As shown in Fig.~\ref{fig:3}a, the measured interference fringes become finer as the loop area increases, indicating an increased force response. The most likely candidate for the reduction of fringe contrast at higher $\Delta q$ is an inhomogeneous transverse field gradient, causing imperfect loop closure \cite{sup}. Fig.~\ref{fig:3}c shows the inverse of the measured fringe period, which is proportional to the force sensitivity, as measured by sinusoidal fits for each interferometer loop. The results match the analytical theory without any fitting parameters. These data clearly demonstrate a well-understood path to scaling up the force sensitivity in continuously-trapped interferometers by increasing loop area.

\section{Programmability}
Beyond this straightforward increase in beamsplitter separation, the inherent programmability and power of Floquet band engineering also offer a wide variety of other methods to enhance or tune the performance of trapped matter-wave interferometers. Possibilities include the use of switchable or pulsed beamsplitters, the inclusion of higher bands beyond $D$, multifrequency Floquet band engineering, and coherent control of the beamsplitting phase. The final set of experiments we describe explores and demonstrates all these capabilities, stocking a versatile toolbox for programmable trapped matter-wave interferometry.

In the initial such experiment, we address a key limitation of the Floquet band engineering approach to matter-wave interferometry: any modulation frequency which resonantly couples two bands (say, $P$ and $D$) at a particular quasimomentum will in general also resonantly couple to other bands at different quasimomenta. Such couplings give rise to undesired ``leaks'' to other states, resulting in parasitic interferometers and reduced contrast. Undesired couplings thus can severely limit the design space for Floquet band interferometers, and become unavoidable for larger-area loops. The problem is illustrated in Fig.~\ref{fig:4}a, which shows the dressed energy band structure of an interferometer subtending about half a Brillouin zone in quasimomentum~\cite{footnote}. The modulation frequency used to create the beamsplitters by coupling the $P$ and $D$ bands at about $0.5$ and $1.5\,\hbar k_{\rm{L}}$ also resonantly couples both the $D$ and $F$ bands and the $P$ and $F$ bands, at different values of quasimomenta within the interferometer loop. While not shown, higher bands beyond $F$ are also coupled in this range via multiphoton transitions. The modulation scheme in Fig.~\ref{fig:4}a shows the simplest way around this problem: smoothly turn on and off the modulation so that the bands are only coupled during the beamsplitting operation. Fig.~\ref{fig:4}b shows a fringe measured in this rather large-area interferometer using pulsed beamsplitters, demonstrating that switchable couplings can be implemented without compromising interferometer performance. This greatly opens up the possible design space for dressed-band interferometry, for instance by allowing loops to extend over multiple Brillouin zones.

Of course, coupling to additional higher bands is not always undesirable; in fact it can be used to design interferometers with enhanced loop area and force sensitivity. Fully flexible use of this capability requires the ability to program arbitrary interband coupling strengths and quasimomenta, which in turn requires the simultaneous use of multiple modulation frequencies. Fig.~\ref{fig:4}c shows the quasienergy band structure (calculated using many-mode Floquet theory~\cite{ho_semiclassical_1983}) of an experiment demonstrating this capability. We modify a loop with $\Delta q = 0.6\,\hbar k_\mathrm{L}$ by applying an additional strong modulation at $196\,\mathrm{kHz}$ to hybridize the $P$ and $F$ bands at quasimomenta $0.85\,\hbar k_\mathrm{L}$ and $1.15\,\hbar k_\mathrm{L}$, with approximately 100\% Landau-Zener transition probability. The loop which includes the $F$ band has a larger area and should be more sensitive to the applied force than the original $P$-$D$ loop. Measured interference fringes for the two interferometers, shown in Fig.~\ref{fig:4}d, demonstrate the predicted decrease in fringe period and increase in sensitivity. This result demonstrates that arbitrary multi-frequency modulations can be used to construct tailored band structures to control and optimize loop geometry and response of trapped matter-wave interferometers.

Finally, to optimize the response of an interferometer to arbitrary perturbations it is helpful to be able to tune the overall phase of the interference fringes. This is straightforwardly accomplished for trapped Floquet-band matter-wave interferometers by adjusting the phase of the lattice modulation waveform which creates one of the beamsplitters. Fig.~\ref{fig:4}e diagrams such an experiment in a $P$-$D$ interferometer with $\Delta q= 0.4\,\hbar k_\mathrm{L}$, which builds on the separated-pulse interferometry demonstrated in Fig.~\ref{fig:4}a-b with the addition of a phase shift between the two beamsplitter modulation pulses. Fig.~\ref{fig:4}f shows the results: for three different overall forces, scanning the phase of the second beamsplitter pulse shifts the fringe by up to $2\pi$. This demonstration of coherent control of beamsplitter phase is useful for determining the contrast when the force cannot be easily scanned, and enables straightforward biasing of interferometer response to the point of maximum sensitivity. Moreover, this method serves as a practical tool for characterizing interferometer stability for a fixed set of parameters. Appendix~\ref{app:tol} shows that the phase-scan fringes are nearly immune to 10\% variations in initial quasimomenta and pulse durations, highlighting the intrinsic robustness inherited from utilizing Landau-Zener transitions for beamsplitters, as achieving the desired transition probability (and thus the maximal contrast) only depends on the pulse overlapping the crossing at some point rather than its exact duration.

\section{Conclusion and Outlook}
We have described and experimentally demonstrated a simple, versatile, and extensible class of trapped matter-wave interferometer constructed from magic Floquet-Bloch band structures in an amplitude-modulated optical lattice. Key virtues of this approach include compactness, flexibility, high ultimate sensitivity unlimited by device size, and intrinsic robustness against trap-induced dephasing and pulse errors. The force sensitivity in the weak-force regime improves as $1/\mathcal{F}$ (Appendix~\ref{app:ext}), a feature that could be exploited in more generality by introducing an accelerated lattice to realize a frame transformation and cancel part of the force. Compact trapped interferometers suitable for weak force measurement may find application in fifth force searches or similar probes of physics beyond the standard model~\cite{RevModPhys.90.025008,Murata_2015,PhysRevD.77.062006,PhysRevLett.124.101101}. 

While our numerical simulations reveal that the force response scales up when the interferometric loop encompasses multiple Bloch oscillations (Appendix~\ref{app:ext}), the ultimate limits on the sensitivity of this technique remain to be explored. Although in the experiments we present the coherence was limited by technical imperfections like transverse field gradient inhomogeneity, this can be straightforwardly improved. Possible avenues for such improvements include using higher-order magic band structures which cancel additional systematics, magnetically insensitive states or isotopes with intrinsically weak interactions~\cite{zhang_trapped-atom_2016}, implementing improved field control, and adding a resonant low-finesse cavity as a mode cleaner. Past work has shown~\cite{fujiwara_transport_2019} that spacetime trajectories can be fully controlled to enable much larger loops, holding at large separations, and executing multiple Bloch oscillations; this means that loop area can in principle grow almost without bound, so any improvements in coherence will directly enhance force sensitivity. The flexibility, power, and large design parameter space offered by the Floquet engineering framework, along with the excellent match between interferometer performance and both numerical and analytical theory, should allow the use of optimal control and machine learning techniques \cite{ledesma_universal_2025} to design more complex interferometer sequences for enhanced robustness and sensitivity. While the low-mass isotope $^7$Li is helpful here for enabling large spatial separation, the technique could be expanded to heavier atoms by using higher-band transitions to generate higher momentum transfer. Finally, given the band-synthesis approach used here, it may be instructive for future developments to compare to, and draw inspiration from, related efforts in driven condensed matter~\cite{basov_towards_2017,oka_floquet_2019}.


\section*{Acknowledgments}

We thank Naceur Gaaloul, Rui Li, and Matthew Glaysher for useful discussion; Samyuktha Ramanan for experimental assistance; and Yifei Bai for physical insight and a critical reading of the manuscript. We acknowledge support from the Army Research Office (W911NF-22-1-0098 and W911NF-20-1-0294), the Noyce Foundation, and the Eddleman Quantum Institute, and from the NSF QLCI program through Grant No. OMA-2016245. E.N.-M. acknowledges support from the UCSB NSF Quantum Foundry through the Q-AMASEi program (Grant No. DMR-1906325). S.N.H. acknowledges support from the NSF NRT program under grant 2152201. 

\section*{Author contributions}

X.C., E.N.-M., X.L., J.L.T., E.Z. and S.N.H. ran the experiment and performed the measurements. X.C. and X.L. implemented lattice and magnetic field stabilization. E.N.-M. performed field curvature cancellation. E.N.-M., J.L.T., E.Q.S., R.S., H.M. and A.C. carried out early experimental explorations. X.C., E.N.-M., X.L., J.L.T. and S.N.H. analyzed the data. X.C. conceptualized the analytical model. X.C., X.L., E.Q.S., E.Z. and A.C. performed numerical calculations. D.M.W. developed the idea for the experiment and supervised the work. X.C., E.N.-M., X.L., J.L.T., E.Z., S.N.H. and D.M.W. wrote the manuscript. All authors contributed to the discussion and interpretation of the results.

\renewcommand{\thesection}{\Alph{section}}
\setcounter{section}{0}

\makeatletter
\renewcommand\section{%
\@startsection
{section}%
{1}%
{\z@}%
{0.8cm \@plus1ex \@minus .2ex}%
{0.5cm}%
{%
\normalfont\small\bfseries
\centering APPENDIX
}%
}%
\def\@seccntformat#1{\csname the#1\endcsname:\quad}
\makeatother

\section{\label{app:exp}Experimental Details}
Our experiments begin with a BEC of $2\times10^5$ $^7$Li atoms produced in a horizontal crossed optical dipole trap with trapping frequencies $(\nu_x,\nu_y,\nu_z)=(151.9,185.0,239.4)\,\si{\hertz}$. The atoms are prepared in the $\ket{f=1, m_f=1}$ ground state, with the magnetic field initially held at \SI{715}{\gauss} along the vertical $z$ direction to ensure a positive scattering length. The field is then ramped down over \SI{1}{\second} to \SI{543.6}{\gauss}, where the $s$-wave scattering length vanishes. Subsequently, the crossed dipole trap is switched off within $\SI{200}{\milli\second}$, while a one-dimensional optical lattice with $\SI{110}{\micro\meter}$ waist is simultaneously ramped up (Fig.~\ref{fig:1}a). This loads the atoms into the ground ($S$) band of the optical lattice with a narrow quasimomentum spread of ${\sim} 0.13\,\hbar k_{\rm{L}}$ (full width at half maximum). Since the lattice is oriented horizontally along the $x$-axis, gravity does not produce a force along the lattice. Instead, a field gradient $\pdv*{B_z}{x}$ generated by the coils is ramped up over $\SI{38.3}{\milli\second}$, which applies a force and initiates Bloch oscillations. The lattice depth is modulated by varying the radio-frequency power sent to an acousto-optic modulator.

\section{\label{app:sp}S-P Transfer}
To perform $P$-$D$ atom interferometry with magic band structures, the atoms must first be prepared in the $P$ band. This is  accomplished via Floquet modulation. After the $\SI{38.3}{\milli\second}$ field gradient ramp, the atoms acquire a momentum such that the central quasimomentum reaches $q=-\hbar k_\mathrm{L}$. A strong lattice amplitude modulation with $\delta V=1.68\,E_\mathrm{R}$ is then ramped up within $\SI{100}{\micro\second}$, opening up a $S$-$P$ hybridized gap at $q\sim 0.5\,\hbar k_\mathrm{L}$. As the matter wave moves towards the Brillouin zone center and traverses the gap, it adiabatically follows the instantaneous Floquet state that connects to the $P$ Bloch band. The modulation is subsequently ramped down within $\SI{100}{\micro\second}$, resulting in a near-unity transfer efficiency to the $P$ band. We chose to couple at this quasimomentum to perform our transfer because the perturbation coupling strength $ \langle\varphi_{S,q}|\cos^2(k_L x)|\varphi_{P,q}\rangle$ is nonexistent at $q=0$ and $q=\hbar k_\mathrm{L}$ due to the parity selection rules. For this same reason, we use the $S$ and $D$ bands for calibration of lattice depth (Supplemental Information) and do not perform a population inversion ($\pi$) at the center of our symmetric $P$-$D$ interferometer.

\section{\label{app:data}Data Acquisition and Analysis}
After band mapping (Fig.~\ref{fig:1}e), the interferometer output appears as spatially separated atomic clouds corresponding to the $P$ and $D$ bands. We perform standard absorption imaging to obtain the atomic column density. The imaging beam is aligned along the $z$-axis, and the $\SI{5}{\micro\second}$ imaging pulse is applied while the magnetic field remains on. To mitigate imaging imperfections, we apply a least-squares regression fringe removal procedure. The $P$ and $D$ band fractions are then calculated by integrating the atom numbers within regions of interest as indicated in Fig.~\ref{fig:1}e.

\section{\label{app:tol}Initial Momentum and Pulse Duration Tolerance}
\begin{figure}[h]
    \centering
    \includegraphics[width=8.5cm]{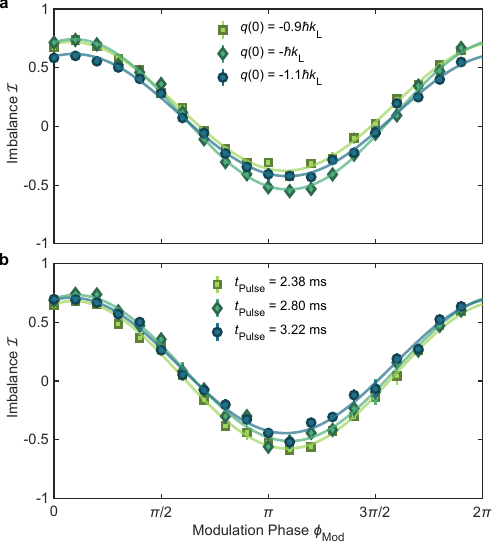}
    \caption{\textbf{Insensitivity to initial momentum and pulse duration}. (\textbf{a}) Output population imbalance as a function of $\phi_\mathrm{Mod}$ for various values of matter-wave quasimomentum $q(0)$ at the beginning of $S$-$P$ modulation. Apart from the varying quasimomentum, the interferometer protocol is identical to Fig.~\ref{fig:4}e. The applied force is fixed at $T_\mathrm{B}=\SI{10.70}{\milli\second}$ and the interferometer loop size is $0.4 \hbar k_\mathrm{L}$. (\textbf{b}) Output population imbalance as a function of $\phi_\mathrm{Mod}$ for various beamsplitter pulse durations. The applied force and loop size are the same as in (a). As in Fig.~\ref{fig:4}e, the two modulation pulses are applied symmetrically across the Brillouin zone edge.} 
    \label{extfig:1}
\end{figure}

Our Floquet-Bloch matter-wave interferometer employs Landau-Zener beamsplitters that provide intrinsic tolerance against fluctuations in both initial quasimomenta $q(0)$ and pulse durations $t_{\rm pulse}$. As long as the remaining duration of the modulation pulse exceeds the Landau-Zener transition timescale when the matter-wave reaches the resonant quasimomentum $q_{\rm r}$, the matter-wave splitting remains faithfully 50-50. Figure.~\ref{extfig:1}a shows the measured interference fringes with $\pm 0.1\hbar k_{\rm L}$ variations of the initial quasimomenta at the beginning of the $S$-$P$ modulation. The resulting fringes exhibits only a 0.06 contrast variation, while the fitted interferometer phases remain nearly unchanged. In Fig.~\ref{extfig:1}b, we instead vary the modulation pulse durations by 15\% and observe a modest 0.025 change of the contrast.

\begin{figure*}[htb]
    \centering
    \includegraphics[width=18cm]{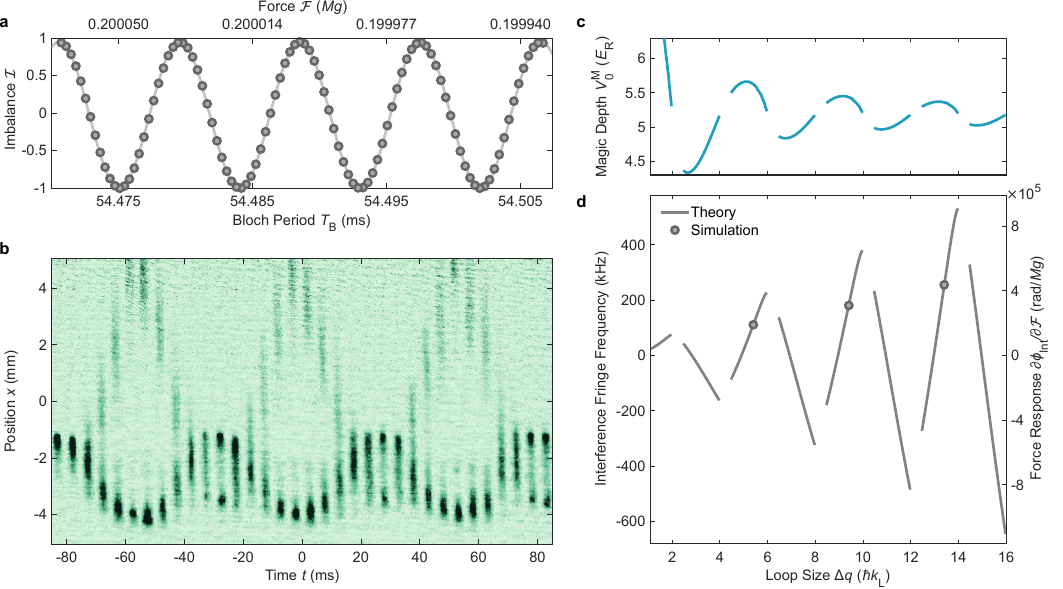}
    \caption{\textbf{Extending the loop size in the weak-force regime}. (\textbf{a}) Numerically simulated interference fringe for loop size $\Delta q=5.4\,\hbar k_{\rm L}$ when $\mathcal{F}\approx 0.2\,Mg$. The solid line is a sinusoidal fit. The lattice depth $V_0 = 5.63\,E_\mathrm{R}$ yields a magic band structure. As in Fig.~\ref{fig:4}a, two modulation pulses with $t_{\rm pulse} = \SI{12.5}{\milli\second}$ are applied symmetrically across the Brillouin zone edge when $q=-1.7\,\hbar k_{\rm L},~3.7\,\hbar k_{\rm L}$. (\textbf{b}) Experimental demonstration of the atoms traversing a $\Delta q = 5.4\,\hbar k_{\rm L}$ $P$-$D$ interferometer loop when $\mathcal{F}\approx 0.2\,Mg$. The lattice depth is kept at $9\,E_{\rm R}$ to prevent $D$ band atoms from moving out of the field of view. (\textbf{c}) Theoretically calculated magic depth as the loop size extends to include multiple Bloch oscillations, assuming the interferometer loop is symmetric about $q=\hbar k_{\rm L}$. The interband coupling due to amplitude modulation vanishes at the Brillouin zone edge and center, so the existence of an upper bound on feasible modulation amplitudes results in the disappearance of a magic structure for a range of loop sizes near integer multiples of $2\hbar k_{\rm L}$ \cite{sup}.  (\textbf{d}) Interferometric response as a function of loop size as defined by the fringe frequency. The sign of the fringe frequency represents the sign of the interferometer phase. Data points represent fit results from numerically simulated fringes, and the solid line is the theoretical prediction obtained from the fit-parameter-free analytical theory. The corresponding force response is calculated at $\mathcal{F} = 0.2\,Mg$.} 
    \label{extfig:2}
\end{figure*}

\section{\label{app:ext}Extending the Loop Size}

Here we present numerical results demonstrating that the force response continues to scale up when the interferometric loop includes multiple Bloch oscillations. For simplicity, we adopt the pulsed $P$-$D$ interferometry scheme, with separation and recombination pulses applied when the quasimomentum reaches $\hbar k_{\rm L} - \Delta q/2$ and $\hbar k_{\rm L} + \Delta q/2$. In large-loop configurations, magic band structures can require smaller lattice depths, leading to increased Landau-Zener tunneling from the $D$ to $F$ band at the Brillouin zone edge. To mitigate this unwanted transition, we assume a reduced force of $\mathcal{F}\approx 0.2\,Mg$ while adjusting the $P$-$D$ modulation strength to maintain 50-50 splitting. Figure~\ref{extfig:2}a shows a simulated interference fringe for $\Delta q = 5.4 \hbar k_{\rm L}$ and $V_0=5.63\, E_{\rm R}$ yielding a magic condition. The top axis indicates a much larger force response compared to the smaller loops in Fig.~\ref{fig:3}. Although large-loop interference has not yet been experimentally observed due to contrast limitations \cite{sup}, the measured  matter-wave trajectory reveals an enlarged spacetime area as the atoms undergo multiple Bloch oscillations (Fig.~\ref{extfig:2}b). 

When $\Delta q$ approaches an integer multiple of $2\hbar k_{\rm L}$, the resonant quasimomenta lie near the Brillouin zone edge or center where the $P$-$D$ coupling vanishes. A magic band structure is absent in such conditions, resulting in the gaps in Fig.~\ref{extfig:2}c and \ref{extfig:2}d. As the loop size increases, the magic depth $V_0^{\rm M}$ oscillates and eventually converges to a constant value (Fig.~\ref{extfig:2}c), which corresponds to the stationary point of the differential dynamical phase integrated over the full Brillouin zone. Figure~\ref{extfig:2}d shows that the force response also oscillates with loop size. At certain loop sizes, destructive interference between the $P$ and $D$ band dynamical phases leads to vanishing force response, which could potentially be combined with  echo techniques for AC force sensing. Despite the oscillations, the overall force response trends up, indicating the expected enhanced sensitivity for longer trapping durations.

\bibliography{interferometry}

\begin{thebibliography}{55}
\bibitem [{foo()}]{footnote}%
  \BibitemOpen
  \href@noop {} {}\bibinfo {note} {Since we vary the modulation amplitude and
  phase and include additional modulation frequencies, the Hamiltonian loses
  perfect discrete time-translation symmetry, so that in principle the Floquet
  formalism is no longer applicable. This is why we use the phrase ``dressed
  energy'' rather than quasienergy here. Nevertheless, the hybridized dressed
  bands generated by a temporally local Floquet picture are of great assistance
  in predicting the dynamics and sensitivity of different modulation
  schemes.}\BibitemShut {Stop}%
\end{thebibliography}
\let\addcontentsline\oldaddcontentsline

\clearpage

\widetext
\begin{center}
    \textbf{\Large Supplemental Material}
\end{center}
\setcounter{figure}{0}
\renewcommand{\figurename}{Fig.}
\makeatletter 
\renewcommand{\thefigure}{S\@arabic\c@figure}
\renewcommand{\theHfigure}{S\@arabic\c@figure}
\makeatother
\renewcommand{\thesection}{\arabic{section}}
\renewcommand{\thesubsection}{\thesection.\arabic{subsection}}
\renewcommand{\thesubsubsection}{\thesubsection.\arabic{subsubsection}}
\setcounter{equation}{0}
\renewcommand{\theequation}{{S}\arabic{equation}}

\makeatletter
\setcounter{section}{0} 
\renewcommand{\thesection}{\arabic{section}} 
\makeatother

\makeatletter
\renewcommand\section{%
\@startsection
{section}%
{1}%
{\z@}%
{0.8cm \@plus1ex \@minus .2ex}%
{0.5cm}%
{%
\normalfont\small\bfseries
\centering
}%
}%
\def\@seccntformat#1{\csname the#1\endcsname.\quad}
\makeatother

\tableofcontents

\clearpage

\section{Floquet-Bloch Atom Interferometry Theory}
In this section, we derive the phase of our Floquet-Bloch atom interferometer from first principles.
\subsection{Bloch Bands in a Static Lattice}
Consider non-interacting atoms confined in a one-dimensional optical lattice.  We assume a periodic boundary condition and a system size $L$ as an integer multiple of the lattice constant, \textit{i}.\textit{e}., $L=N\lambda/2$ where $N$ is a large integer and $\lambda$ is the lattice wavelength.  The single-particle Hamiltonian is given by
\begin{align}
    \hat{H}_0=\frac{\hat{p}^2}{2M} 
    - V_0\cos^2{\left(k_\mathrm{L} \hat{x}\right)},
    \label{eq:atom}
\end{align}
where $M$ is the atomic mass; $V_0$ is the lattice depth; $k_{\mathrm{L}} = 2\pi / \lambda$ is the lattice laser angular wavenumber; $\hat{x}$ and $\hat{p}$ are position and momentum operators, respectively. Leveraging the discrete translational symmetry
\begin{align}
    \hat{H}_0\left(\hat{x} + \lambda / 2\right) = \hat{H}_0\left(\hat{x}\right),
\end{align}
Bloch's theorem gives rise to the eigenstates $\ket{\varphi_{n,q}}$, where $n=0,1,2...$ denote Bloch band indices (equivalent to $S,P,D...$); $q = 2\pi \hbar j / L$ is the quasimomentum; $j = -N / 2 +1, ... , 0 , ... ,N / 2$; and  $\hbar$ is the reduced Planck constant. In position space, the eigenstates satisfy
\begin{align}
    \varphi_{n,q}(x) & = \braket{x}{\varphi_{n,q}}=
    \frac{1}{\sqrt{N}} u_{n,q}(x) 
    \mathrm{e}^{\mathrm{i} q x / \hbar}, \nonumber \\
    u_{n,q}(x) &= u_{n,q}(x+\lambda / 2).
    \label{eq:ansatz}
\end{align}
Substituting this ansatz into the time-independent Schr\"odinger equation results in an eigenvalue problem for the Bloch function $u_{n,q}(x)$:
\begin{align} 
    \left[
    \frac{(-\mathrm{i} \hbar \partial_x + q)^2}{2M} 
    - V_0 \cos^2{(k_\mathrm{L} {x})}
    \right]
    u_{n,q}(x) = 
    E_{n,q} u_{n,q}(x).
    \label{eq:eig}
\end{align}
This eigenvalue problem can be solved numerically to extract both $u_{n,q}(x)$ and the band energy $E_{n,q}$ (Fig.~\ref{sifig:1}a). Notice that if we impose the normalization of $u_{n,q}(x)$ as 
\begin{align}
    \label{eq:unrom}
    \int_{-\lambda / 4}^{\lambda / 4} 
    \dd{x} u^*_{n,q}(x) u_{n',q}(x) = \delta_{n,n'},
\end{align}  
the Bloch states are orthonormal:
\begin{align}
    \braket*{\varphi_{n,q}}{\varphi_{n',q'}} = \int_{-L / 2}^{L / 2} \dd{x}
    \varphi_{n,q}^{*}(x) \varphi_{n',q'}(x) = \delta_{n,n'}\delta_{q,q'}.
\end{align}
In the following discussion, we take $q$ as a continuous variable, which only makes sense in the $L\rightarrow \infty$ limit.

\subsection{Floquet-Bloch Bands in a Driven Lattice}
Next, we consider a continuous and sinusoidal amplitude-modulation of the lattice,
\begin{align}
    \hat{H}_\mathrm{Mod}(t) = \frac{\hat{p}^2}{2M} - \left( V_0+\delta V\cos\omega t \right) \cos^2\left(k_\mathrm{L} \hat{x}\right),
    \label{eq:hs}
\end{align}
where $\delta V$ and $\omega$ are the modulation depth and frequency, respectively. This Hamiltonian exhibits a two-fold (spatial and temporal) translational symmetry that gives rise to the Floquet-Bloch states $\ket{\tilde{\varphi}_{l,q}(t)}$ with wavefunctions given by
\begin{align}
    \tilde{\varphi}_{l,q}(x,t) = \braket{x}{\tilde{\varphi}_{l,q}(t)} = 
    \exp[\frac{\mathrm{i}}{\hbar}(qx-\tilde{E}_{l,q}t)]\tilde{u}_{l,q}(x,t),
    \label{eq:fb}
\end{align}
where $l$ is the Floquet-Bloch band index, $\tilde{E}_{l,q}$ is the quasienergy, and $\tilde{u}_{l,q}(x,t)$ is the Floquet-Bloch function that satisfies 
\begin{align}
    \tilde{u}_{l,q}(x,t) = \tilde{u}_{l,q}(x+\lambda/2,t) = \tilde{u}_{l,q}(x,t+2\pi/\omega).
\end{align}
Both $\tilde{E}_{l,q}$ and $\tilde{u}_{l,q}(x,t)$ can be obtained from numerically diagonalizing the Floquet Hamiltonian $\hat{H}_\mathrm{F}$, defined as
\begin{align}
    \exp(-\frac{2\pi\mathrm{i}\hat{H}_\mathrm{F}}{\hbar \omega}) = \mathcal{T}\exp \left\{-\frac{\mathrm{i}}{\hbar}\int_{0}^{2\pi/\omega} \dd{t} \hat{H}_\mathrm{Mod}(t) \right\},
    \label{eq:FH}
\end{align}
where $\mathcal{T}$ is the time-ordering operator.

\begin{figure}[ht]
    \centering
    \includegraphics[width=18cm]{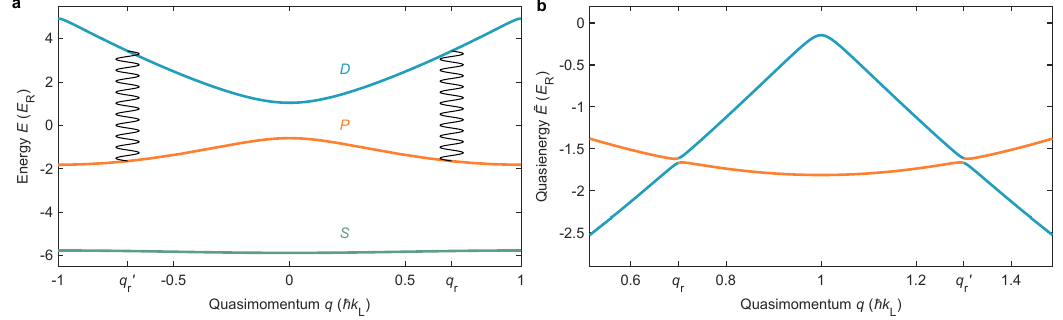}
    \caption{\textbf{Band structure}. (\textbf{a}) The three lowest bands of a static optical lattice, labeled $S$ (green), $P$ (orange) and $D$ (blue). Amplitude modulation couples the latter two bands at the points of resonance $q_\mathrm{r}, q_\mathrm{r}'$ found by Eq.~(\ref{eqn:modulation_resonance_condition}), indicated by the connecting ``drive photons.'' (\textbf{b}) Floquet-Bloch bands resulting from the amplitude modulation in (a) hybridizing the $P$ and $D$ bands. Orange and blue indicate the dominant character of the band relative to the original static bands.}
    \label{sifig:1}
\end{figure}

It is shown in Fig.~\ref{sifig:1}b that a resonant modulation opens up gaps between two Floquet-Bloch bands. The modulation frequency $\omega$ determines the coupled Bloch band indices $(n,n')$ and the resonant quasimomentum $q_\mathrm{r}$, based on the resonance condition 
\begin{align}
    E_{n,q_{\mathrm{r}}}-E_{n',q_{\mathrm{r}}} = \hbar \omega.
    \label{eqn:modulation_resonance_condition}
\end{align}
Meanwhile, the modulation depth $\delta V$ and the off-diagonal ($n\ne n'$) coupling matrix element govern the size of the gap $\Delta_{q_\mathrm{r}}$ at the avoided crossings:
\begin{equation}
    \Delta_{q_\mathrm{r}} \approx \delta V \abs{\matrixel{\varphi_{n,q_\mathrm{r}}}{\cos^2\left(k_\mathrm{L}\hat{x}\right)}{\varphi_{n',q_\mathrm{r}}}}.
    \label{eqn:band_gap}
\end{equation}

\subsection{Interferometer Phases}

When a uniform force $\mathcal{F}$ is applied along the lattice direction, the Hamiltonian reads
\begin{align}
    \hat{H}(t) = \frac{\hat{p}^2}{2M} - \left( V_0+\delta V\cos\omega t \right) \cos^2\left(k_\mathrm{L} \hat{x}\right) - \mathcal{F}\hat{x}.
    \label{eq:h0sup}
\end{align}
Applying a gauge transformation $\hat{U} = \exp(-\mathrm{i} \mathcal{F}t\hat{x}/\hbar)$, we recover the spatial periodicity,
\begin{align}
    \hat{H} \rightarrow \hat{H}' 
    &= \hat{U}\hat{H}\hat{U}^{\dagger}  + \mathrm{i}\hbar \left(\partial_t \hat{U}\right) 
    \hat{U}^{\dagger} \nonumber \\
    &= \frac{(\hat{p}+\mathcal{F}t)^2}{2M} - \left( V_0+\delta V\cos\omega t \right) \cos^2\left(k_\mathrm{L} \hat{x}\right).
    \label{eq:hsup}
\end{align}
Henceforth, we omit the prime on $\hat{H}'$ and assume $\hat{U}$ is always applied.

\begin{figure}[ht]
    \centering
    \includegraphics[width=18cm]{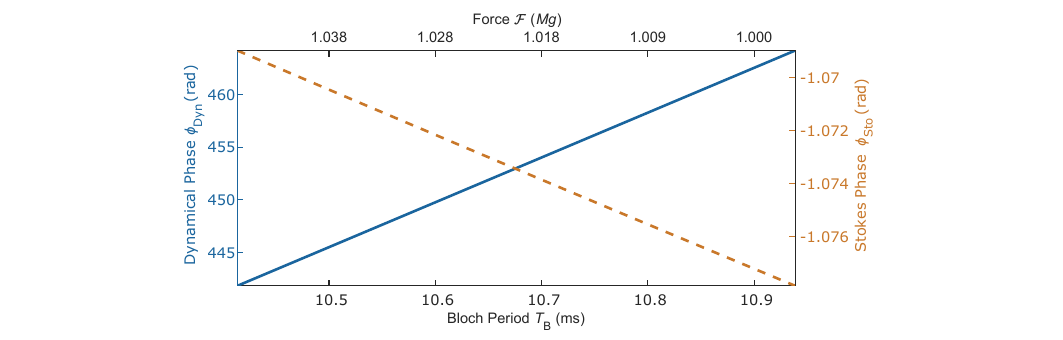}
    \caption{\textbf{Interferometer Phases}. The dynamical (blue, solid) and Stokes (brown, dashed) phases as a function of Bloch period. The lattice and modulation parameters are the same as in Fig.~\ref{fig:1} in the main text. Note the very different $y$ axes for the different phases; the change in the Stokes phase is generally negligible.}
    \label{sifig:2}
\end{figure}

We define $\mathcal{Q}=\mathcal{F}t$ as the slowly varying parameter in the Hamiltonian (\ref{eq:hsup}) that facilitates the instantaneous-Floquet-state (IFS) formalism \cite{ikeda_floquet-landau-zener_2022}.
Since in our experiments the Bloch frequency $\omega_{\rm B} = \mathcal{F}\lambda/2\hbar \ll \omega$, this is a valid separation of time scales. A set of IFSs $\ket*{\tilde{\zeta}_{l,q}(t)}_{\mathcal{Q}}$ can be given by diagonalizing the Hamiltonian (\ref{eq:hsup}) while fixing $\mathcal{Q}$:
\begin{align}
    \ket*{\tilde{\zeta}_{l,q}(t)}_{\mathcal{Q}} = \ket{{\tilde{\varphi}_{l,q+\mathcal{Q}}(t)}}.
\end{align}
Intuitively, one would imagine that the quantum state can adiabatically follow the IFS when the control parameter $\mathcal{Q}$ varies slowly. However, an exact Floquet adiabatic limit is likely absent due to the dense quasienergy spectrum and infinitely many avoided-crossings from multi-Floquet-photon resonances \cite{hone_time-dependent_1997}. We adopt a coarse-graining argument \cite{holthaus_floquet_2016}, which states that an infinite number of avoided-crossings below a certain scale can be ignored because of the finite experimental time scale. Away from the sizable avoided-crossings, the quantum state $\ket{\psi(t)}$ still adiabatically follows the IFS in an approximate sense,
\begin{align}
    \ket{\psi(t)} \approx \mathrm{e}^{-\mathrm{i}\phi_{\mathrm{Dyn},l}(t)}\ket*{{\tilde{\zeta}_{l,q_0}(t)}}_{\mathcal{Q}}
    = \mathrm{e}^{-\mathrm{i}\phi_{\mathrm{Dyn},l}(t)}\ket{{\tilde{\varphi}_{l,q_0+\mathcal{Q}}(t)}},~
    \mathrm{if}~\ket{\psi(0)} = \ket{\tilde{\varphi}_{l,q_0}(0)},
    \label{eq:adiabaticsup}
\end{align}
where 
\begin{equation}
    \phi_{\mathrm{Dyn},l}(t)=\frac{1}{\hbar \mathcal{F}}\int_{q_0}^{q_0 + \mathcal{F}t} \dd{q} \tilde{E}_{l,q}
\end{equation} 
is the dynamical phase, and $q_0$ is the initial quasimomentum. Combined with the periodicity of quasimomentum, Eq.~(\ref{eq:adiabaticsup}) confirms the presence of Bloch oscillations in Floquet-Bloch bands. To harness the force sensitivity of $\phi_{\mathrm{Dyn},l}(t)$, we introduce interference between two Floquet-Bloch bands via Landau-Zener tunneling \cite{landau1932theorie1,landau1932theorie2,zener_non-adiabatic_1932} at avoided-crossings. This occurs when the Bloch frequency $\omega_\mathrm{B}$ is comparable to the gap size $\Delta_{q_\mathrm{r}}/\hbar$. As illustrated in Fig.~\ref{fig:1}c in the main text, when a Floquet-Bloch atom in the upper band passes the avoided crossing at $q=q_\mathrm{r}$, it splits into a superposition of two Floquet-Bloch waves, and the ratio can be adjusted by tuning $\delta V$, and thus the gap size, to realize a 50-50 beam splitter. The two waves accumulate distinct dynamical phases before recombining at the second avoided crossing 
at $q=q'_\mathrm{r}$. This setup forms a Landau–Zener–St\"uckelberg–Majorana interferometer \cite{shevchenko_landauzenerstuckelberg_2010,ivakhnenko_nonadiabatic_2023}, whose output is measured as the final population imbalance between the Floquet-Bloch bands given by the adiabatic-impulse approximation \cite{ikeda_floquet-landau-zener_2022}
\begin{align}
    \mathcal{I} = p_{\rm L} - p_{\rm U} \approx 4p(1-p)\cos\phi_\mathrm{Int},
\end{align}
where $\rm L(U)$ denotes the lower (upper) Floquet-Bloch band and $p_{\rm L(U)}$ represents the band population; 
\begin{align}
    p = \mathrm{e}^{-2\pi\delta}
\end{align}
is the Landau-Zener transition probability;
\begin{align}
    \delta=\frac{\Delta_{q_\mathrm{r}}^2}{4\hbar v}
\end{align}
is the adiabaticity parameter; 
\begin{align}
    v= \mathcal{F}\sqrt{2\Delta_{q_\mathrm{r}}\abs{\pdv[2]{\tilde{E}_{\mathrm{L},q_\mathrm{r}}}{q}}}
\end{align}
is the Landau-Zener sweep velocity;
\begin{align}
    \phi_\mathrm{Int} = \phi_\mathrm{Dyn}+2\phi_\mathrm{Sto}
\end{align}
is the interferometer phase; 
\begin{align}
    \phi_\mathrm{Sto}=-\frac{\pi}{4}+\delta(\ln\delta-1)+\arg\Gamma(1-\mathrm{i}\delta)
\end{align}
is the Stokes phase \cite{shevchenko_landauzenerstuckelberg_2010,ivakhnenko_nonadiabatic_2023},
and 
\begin{align}
    \phi_\mathrm{Dyn} = \frac{1}{\hbar \mathcal{F}}\int_{q_\mathrm{r}}^{q'_\mathrm{r}} \dd{q} \left( \tilde{E}_{\mathrm{U},q} - \tilde{E}_{\mathrm{L},q}\right)
\end{align}
is the (differential) dynamical phase which is \textit{inversely} proportional to the applied force. Since the Stokes phase $\phi_\mathrm{Sto}$ only weakly depends on $\mathcal{F}$ (Fig.~\ref{sifig:2}), the force sensitivity of this interferometer is dominated by the dynamical phase $\phi_\mathrm{Dyn}$, which scales with the energy-momentum area enclosed by the interferometer loop.

\subsection{Wave Packets}
The above analysis only applies to a Floquet-Bloch wave with a single quasimomentum component, which is spatially delocalized. Realistically, the quantum state of the atomic cloud in experiments is a localized wave packet, which justifies the following initial condition as a superposition of quasimomentum states,
\begin{align}
    \ket{\psi(0)} = \int_{-\hbar k_{\rm L}}^{\hbar k_{\rm L}}
    \dd{q_0} \mathcal{P}(q_0) \ket{\tilde{\varphi}_{l,q_0}(0)},
\end{align}
where $\mathcal{P}(q_0)$ is the quasimomentum distribution function. If the quasimomentum distribution is narrow enough, all the quasimomentum components can traverse and complete the interferometric loop, thanks to the robustness of Landau-Zener transitions. In such conditions, the interferometer output, measured as the $P$ and $D$ band population imbalance, is not different from the output with a single quasimomentum component.

\subsection{Geometric Phase}
Since the atoms follow an adiabatic path along the IFSs, one may be concerned with the appearance of a geometric phase in the calculation of the interferometer phase. In particular, the analysis of a topologically invariant Zak phase in \cite{zak_berrys_1989} only applies to static lattices and paths across the entire Brillouin zone, while our considerations require a modulated lattice across part of the Brillouin zone. It is shown in \cite{breuer_quantum_1989} that the geometric phase for periodically driven systems in the extended Hilbert space takes the form
\begin{align}
    \gamma_l(C) = \frac{1}{T} \int_C \dd{q} \int_0^T \dd{t} \langle \tilde{u}_{l,q}(t) | \mathrm{i}\partial_q \tilde{u}_{l,q}(t)\rangle,
    \label{eqn:berry}
\end{align}
where $T=2\pi/\omega$ is the period of the drive; the states $\langle x|\tilde{u}_{l,q}(t)\rangle=\tilde{u}_{l,q}(x,t)$; and $C$ is the path taken through the parameter space, which for our interferometric loop consists of the portion of the Brillouin zone bounded by $q\in[q_\mathrm{r}, q_\mathrm{r}']$. Then, the differential geometric phase for our two-band interferometer is given by:
\begin{align}
    \gamma_\mathrm{U} - \gamma_\mathrm{L} = \int_{q_\mathrm{r}}^{q'_\mathrm{r}} \dd{q} \left[\tilde{X}_\mathrm{UU}(q) - \tilde{X}_\mathrm{LL}(q)\right],
\end{align}
where the modified Berry connection is given by (\ref{eqn:berry}):
\begin{align}
    \tilde{X}_{ll}(q) = \frac 1T \int_0^T \dd{t} \int_{-\lambda/4}^{\lambda /4} \dd{x} \tilde{u}_{l,q}^*(x,t)\left[ \mathrm{i} \frac{\partial }{\partial q} \tilde{u}_{l,q}(x,t)\right].
    \label{eqn:connection}
\end{align}
Numerical calculations of the Zak phase (evaluated along paths spanning the entire Brillouin zone) for various lattice and modulation parameters reveal that the Floquet-Bloch band is topologically trivial ($\gamma = 2 \pi \mathcal{N}$ with $\mathcal{N} \in \mathbb{Z}$), so under a time-independent gauge transformation $\tilde{u}_{l,q}(x,t) \rightarrow \tilde{u}'_{l,q}(x,t) =e^{\mathrm{i}\tilde{\phi}_l(q)} \tilde{u}_{l,q}(x,t)$, the connection transforms as
\begin{align}
    \tilde{X}_{ll}\longrightarrow\tilde{X}'_{ll} = \tilde{X}_{ll} - \frac 1T \int_0^T \dd{t} \frac{\partial \tilde{\phi}_l}{\partial q} = \tilde{X}_{ll} - \frac{\partial \tilde{\phi}_l}{\partial q},
\end{align}
and likewise the differential geometric phase as
\begin{align}
    \gamma_\mathrm{U} - \gamma_\mathrm{L} = \int_{q_\mathrm{r}}^{q'_\mathrm{r}} \dd{q} \left[\tilde{X}'_\mathrm{UU}(q) - \tilde{X}'_\mathrm{LL}(q)\right].
\end{align}
A judicious choice of $\tilde{\phi}_l$ can thus be used to gauge away the differential geometric phase by solving the following equation for both the upper and lower bands:
\begin{align}
    \gamma_l = \int_{q_\mathrm{r}}^{q'_\mathrm{r}} \dd{q}  \tilde{X}'_{ll}(q) = \int_{q_\mathrm{r}}^{q'_\mathrm{r}} \dd{q} \left(\tilde{X}_{ll} - \frac{\partial \tilde{\phi}_l}{\partial q} \right) = 0.
\end{align}
Substitution of Equation (\ref{eqn:connection}) for $\tilde{X}_{ll}$ yields:
\begin{align}
    \gamma_l = \frac 1T \int_0^T \dd{t} \int_{q_\mathrm{r}}^{q'_\mathrm{r}} \dd{q}\int_{-\lambda/4}^{\lambda /4} \dd{x} \tilde{u}_{l,q}^*(x,t)\left[\mathrm{i} \frac{\partial }{\partial q} \tilde{u}_{l,q}(x,t)\right] -  \left[\tilde{\phi}_l\left(q'_\mathrm{r}\right) - \tilde{\phi}_l\left(q_\mathrm{r}\right)\right] = 0.
\end{align}
A solution for $\tilde{\phi}_l$ is then obtained:
\begin{align}
    \tilde{\phi}_l\left(q'_\mathrm{r}\right) - \tilde{\phi}_l\left(q_\mathrm{r}\right) = \frac 1T \int_0^T \dd{t}\int_{q_\mathrm{r}}^{q'_\mathrm{r}} \dd{q}\int_{-\lambda/4}^{\lambda /4} \dd{x} \tilde{u}_{l,q}^*(x,t)\left[\mathrm{i} \frac{\partial }{\partial q} \tilde{u}_{l,q}(x,t)\right].
    \label{eqn:gauge}
\end{align}
Therefore, the geometric phase across an arbitrary portion of the Brillouin zone is pure gauge, so it is not physically observable and we can ignore it in our calculations of the interferometer phase. 

\subsection{Dressed State Picture}
When the amplitude modulation is pulsed, ramped, or involves multiple frequency components, there is formally no exact discrete time-translation symmetry; therefore, the Floquet formalism is not perfectly applicable. To address this, we introduce a dressed state approach, which provides an alternative framework for understanding the operation of the interferometer.

Consider the following Hamiltonian where the amplitude modulation has an envelope $\alpha(t)$:
\begin{align}
    \hat{H}(t) = \frac{\hat{p}^2}{2M} - V_0\left[ 1+\alpha(t)\cos\omega t \right] \cos^2\left(k_\mathrm{L} \hat{x}\right) - \mathcal{F}\hat{x}.
    \label{eq:h0general}
\end{align}
After a gauge transformation $\hat{U} = \exp(-\mathrm{i} \mathcal{F}t\hat{x}/\hbar)$, the Hamiltonian recovers discrete space-translation symmetry, 
\begin{align}
    \hat{H}(t) \rightarrow  
    \hat{U}\hat{H}\hat{U}^{\dagger}  + \mathrm{i}\hbar \left(\partial_t \hat{U}\right) 
    \hat{U}^{\dagger} = \frac{(\hat{p}+\mathcal{F}t)^2}{2M} - V_0\left[ 1+\alpha(t)\cos\omega t \right] \cos^2\left(k_\mathrm{L} \hat{x}\right).
    \label{eq:hgeneral}
\end{align}
We assume initially the wave function has the form of a Bloch wave under this gauge,
\begin{align}
    \psi(x,t=0)=\braket*{x}{{\psi}(t=0)} = \mathcal{U}(x,t=0) \mathrm{e}^{\mathrm{i} q_0 x / \hbar},
\end{align}
where $\mathcal{U}(x,t=0)$ is periodic in space with periodicity of $\lambda / 2$ and $q_0$ is the initial quasimomentum. Because of the discrete space-translation symmetry of (\ref{eq:hgeneral}), the quasimomentum is conserved and we can write, at a later time $t$, 
\begin{align}
    \psi(x,t)=\braket*{x}{{\psi}(t)} = \mathcal{U}(x,t) \mathrm{e}^{\mathrm{i} q_0 x / \hbar},
\end{align}
where $\mathcal{U}(x,t)$ is also periodic in space. Performing a backward gauge transformation $\hat{U}^{\dagger} = \exp(\mathrm{i} \mathcal{F}t\hat{x}/\hbar)$, we have
\begin{align}
    \psi(x,t)=\braket*{x}{{\psi}(t)} \rightarrow \psi'(x,t)=\braket*{x}{{\hat{U}^{\dagger}\psi}(t)} = \mathcal{U}(x,t) \mathrm{e}^{\mathrm{i} (q_0 +\mathcal{F}t)x/\hbar}.
\end{align}
Therefore, in the original gauge the time-dependent wavefunction can be expressed as a superposition of Bloch states with the instantaneous quasimomentum $q=q_0+\mathcal{F}t$ (omitting the prime symbol for simplicity):
\begin{align}
    \ket{\psi(t)} = \sum_{n=0}^{\infty} c_n(t)\ket{\varphi_{n,q_0+\mathcal{F}t}}.
\end{align}
In the following, we assume the summation can be truncated to include only two Bloch bands, e.g., the $P$ and $D$ bands:
\begin{align}
    \ket{\psi(t)} \approx \sum_{n=1}^{2} c_n(t)\ket{\varphi_{n,q_0+\mathcal{F}t}}.
    \label{eq:2band}
\end{align}
This two-band approximation is valid when we  consider dynamics near the $P$-$D$ resonance driven by the modulation, or when the modulation is weak enough such that it barely couples transitions outside of the $P$-$D$ subspace. Substituting Eq.~(\ref{eq:2band}) into the time-dependent Schr\"odinger equation with the Hamiltonian (\ref{eq:h0general}), we have 
\begin{align}
    \mathrm{i} \hbar \pdv{t} \mqty({c}_{2} \\ {c}_{1}) 
    = {\bm{\mathrm{H}}}_{\mathrm{Eff}}\mqty({c}_{2} \\ 
    {c}_{1}), ~ 
    {\bm{\mathrm{H}}}_{\mathrm{Eff}}(t) 
     \approx \mqty(E_{2,q_0+\mathcal{F}t}- \hbar \omega 
    & -\hbar\Omega(t) / 2 \\ 
    -\hbar\Omega^{*}(t) / 2
    & E_{1,q_0+\mathcal{F}t}),
    \label{eq:heff}
\end{align}
where the Rabi frequency is $\Omega(t) = {\alpha(t)V_0 \matrixel{\varphi_{1,q_0+\mathcal{F}t}}{\cos^2\left(k_\mathrm{L}\hat{x}\right)}{\varphi_{2,q_0+\mathcal{F}t}}}/{\hbar}$. To derive the effective Hamiltonian ${\bm{\mathrm{H}}}_{\mathrm{Eff}}(t)$, we have applied a rotating-frame transformation and used the rotating-wave approximation (RWA), while neglecting the linear potential term $-\mathcal{F}\hat{x}$, which typically contributes only a small perturbation for $P$ and $D$ bands. 

Eq.~(\ref{eq:heff}) implies that the system's dynamics can be understood as those of a two-level system with time-dependent parameters. In fact, the theory of St\"uckelberg interferometers \cite{shevchenko_landauzenerstuckelberg_2010,ivakhnenko_nonadiabatic_2023} can be directly applied for such systems. The interferometer loop is formed by two Landau-Zener transitions between the dressed states, and the interferometer phase is dominated by the dynamical phase, given as
\begin{align}
    \phi_\mathrm{Dyn} = \frac{1}{\hbar }\int_{t_\mathrm{r}}^{t'_\mathrm{r}} \dd{t} \left[ {\epsilon}_{{+}}(t) - {\epsilon}_{{-}}(t)\right],
    \label{eq:dyn2}
\end{align}
where $ {\epsilon}_{\pm}(t)$ are the dressed energies obtained by diagonalizing the effective Hamiltonian ${\bm{\mathrm{H}}}_{\mathrm{Eff}}(t)$ at each moment in time, and $t_\mathrm{r},t'_\mathrm{r}$ correspond to solutions of the resonance condition $E_{2,q_0+\mathcal{F}t}- \hbar \omega = E_{1,q_0+\mathcal{F}t}$. This framework provides an alternative method for computing the interferometer phase that does not rely on exact discrete time-translational symmetry. Moreover, this dressed-state approach can be straightforwardly extended to include multi-tone driving, multi-band dynamics, and varying modulation phase.

\section{Magic Depth Calculation}

Since the Stokes phase is monotonic in $\delta$ and bounded by $[-\pi/2, -\pi/4]$, it varies far less than the dynamical phase for the same range of force and lattice depth. Thus, while exploring the parameter space of $V_0, q_\mathrm{r},$ and $\omega$, we approximate the magic condition as
\begin{align}
    0 &=\frac{\partial \phi_\mathrm{Int}}{\partial V_0} \nonumber \\ 
    &=\frac{\partial\phi_\mathrm{Dyn}}{\partial V_0} + 2\frac{\partial\phi_\mathrm{Sto}}{\partial V_0} \nonumber \\ 
    &\approx\frac{\partial\phi_\mathrm{Dyn}}{\partial V_0}.
\end{align}
We focus on interferometer loops that couple the $P$ and $D$ bands at two quasimomenta $q_{\rm r},~q_{\rm r}'$ which are symmetric about the Brillouin zone edge. In such cases, when the loop size $\Delta q$ is an integer multiple of $2\hbar k_{\rm L}$, the resonant quasimomenta $q_{\rm r},~q_{\rm r}'$ fall at the Brillouin zone edge or center, where the modulation coupling strength vanishes. As a result, interferometer loops cannot form when $\Delta q \simeq 2\mathcal{N}\hbar k_{\rm L}$ with $\mathcal{N}\in \mathbb{Z}^+$, leading to the gaps in Fig.~\ref{extfig:2}c and \ref{extfig:2}d. As indicated in the main text, we have to pulse the modulation to avoid unwanted band couplings when the loop size becomes larger, which formally makes the dressed state approximation necessary. To evaluate the magic condition using dressed energies, we adopt the following two-step protocol.

First, since our amplitude modulation is perturbatively weak, static Bloch band energies well approximate the actual Floquet-Bloch band quasienergies. We use numerical integration of these static band energies across the desired quasimomentum range $[q_\mathrm{r},q_\mathrm{r}']$ with sampling resolution $10^3$ and numerical differentiation with respect to lattice depth $V_0$ to estimate the magic condition and the modulation frequency/depth required to couple the $P$ and $D$ bands at the desired quasimomenta. 

We then confirm and refine this prediction with a more sophisticated calculation using Eq.~(\ref{eq:dyn2}) at a fixed modulation frequency and modulation depth (rather than fixed resonant quasimomentum), as in Fig.~\ref{fig:2}a in the main text. Since both Bloch and Floquet-Bloch bands become flat in the limit of infinite lattice depth, the differential dynamical phase between Floquet-Bloch bands vanishes in the same limit for fixed $q_\mathrm{r}$, precluding the possibility of a local minimum in $\phi_\mathrm{Dyn}$. We instead allow $q_\mathrm{r}$ to vary for fixed modulation frequency; consequently, the $\phi_\mathrm{Dyn}$ curves terminate at some minimum and/or maximum lattice depth beyond which the bands are no longer resonant with the modulation for any quasimomentum.

Likewise, numerical calculation of $\phi_\mathrm{Int}$ as a function of $V_0$ and $\Delta q$ allows estimation of the lattice tolerance in the main text Fig. \ref{fig:3}b, defined as the deviation from the magic depth $V_0^\mathrm{M}$ resulting in $\phi_\mathrm{Int}^\mathrm{M}\pm\pi/4$.

\section{Numerical Calculation}

In this section, we describe numerical methods for obtaining (Floquet-) Bloch energy bands. We decompose the Bloch function (\ref{eq:ansatz}) using Fourier modes:
\begin{align}
    u_{n,q}(x)= \sum_{j=-\infty}^{\infty}c_{n,q}^{(j)} \exp \left(  2\mathrm{i} jk_\mathrm{L}x\right).
\end{align}
Substituting this ansatz into the Schr\"odinger equation (\ref{eq:eig}) gives the eigenvalue equation
\begin{align}
    \sum_{j'=-{\infty}}^{{\infty}} \left[ \left( \frac{(2j\hbar k_\mathrm{L} + q)^2}{2M}  - \frac{V_0}{2}\right) \delta_{j,j'} - \frac{V_0}{4} (\delta_{j-1,j'} + \delta_{j+1, j'}) \right] c_{n,q}^{(j')} = E_{n,q} c_{n,q}^{(j)},
\end{align}
which can be solved numerically if we truncate the Fourier components up to appropriate limits $\pm j_{\rm Max}$. Using a similar approach for the Floquet-Bloch function (\ref{eq:fb})
\begin{align}
    \tilde{u}_{l,q}(x,t) = \sum_{j,k=-\infty}^{\infty} c_{l,q}^{(j,k)} \exp (\mathrm{i}(2jk_\mathrm{L}x - k\omega t)),
\end{align}
results in the eigenvalue equation:
\begin{equation}
    \begin{split}
       \tilde{E}_{l,q}c_{l,q}^{(j,k)}=\sum_{j',k'=-\infty}^{\infty}c_{l,q}^{(j',k')}\biggl\{& \left( \frac{(2j\hbar k_\mathrm{L}+q)^2}{2M}-\frac{V_0}{2}-k\hbar\omega\right)\delta_{j,j'}\delta_{k,k'}\\ 
        &- \frac{V_0}{4}\left(\delta_{j-1,j'}\delta_{k,k'}+\delta_{j+1, j'}\delta_{k,k'}\right)\\
        &- \frac{\delta V}{4}\left(\delta_{j,j'}\delta_{k+1,k'}+\delta_{j, j'}\delta_{k-1,k'}\right) \\ 
        &- \frac{\delta V}{8}\left(\delta_{j-1,j'}\delta_{k+1,k'}+\delta_{j+1, j'}\delta_{k+1,k'}\right)\\
        &- \frac{\delta V}{8}\left(\delta_{j-1,j'}\delta_{k-1,k'}+\delta_{j+1, j'}\delta_{k-1,k'}\right) \biggr\}.
    \end{split}
    \label{eqn:effective_Floquet}
\end{equation}
In practice, $c_{l,q}^{(j,k)}$ and the $j,j',k,k'$ indices can be flattened to produce a vector eigenvalue equation where the eigenvalues are the quasienergies. Again, this eigenvalue problem can be solved if we truncate the Fourier components in space and time up to appropriate limits $\pm j_{\rm Max}$ and $\pm k_{\rm Max}$. This method is used along with diagonalizing the Floquet Hamiltonian in (\ref{eq:FH}), and both approaches produce the same numerical results.


\section{Calibration}

\subsection{Lattice Depth}
\begin{figure}[ht]
    \centering
    \includegraphics[width=18cm]{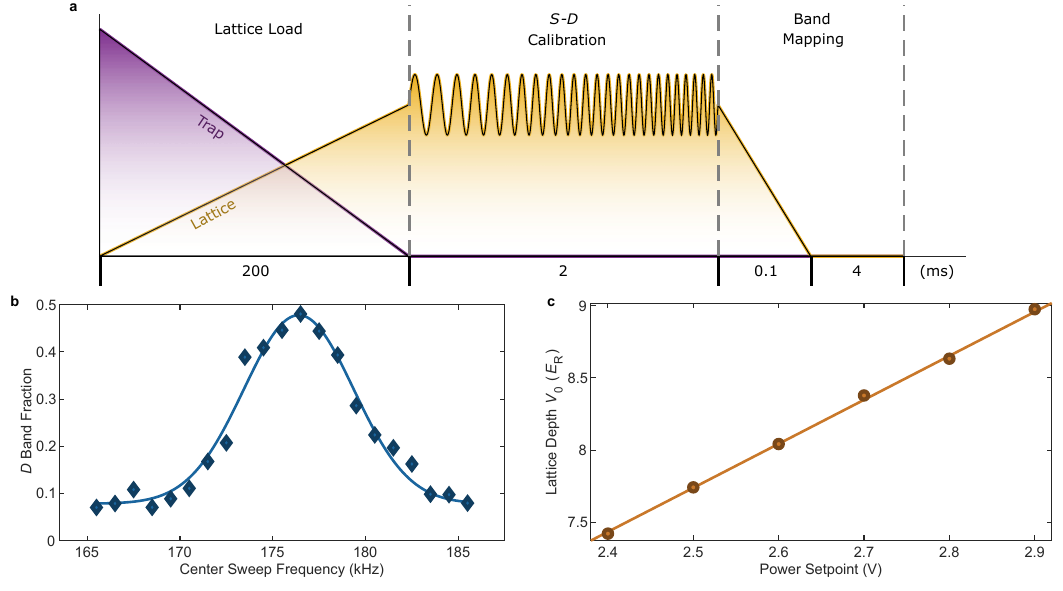}
    \caption{\textbf{Lattice depth calibration.} (\textbf{a}) Experimental sequence for lattice depth calibration. The dipole trap is ramped down and the lattice is ramped up over $200\,\mathrm{ms}$, followed by a resonant amplitude modulation with a $\SI{1}{\kilo\hertz}$ frequency sweep over $2\,\mathrm{ms}$. Band-mapping separates the $S$ and $D$ band populations for detection. (\textbf{b}) $D$ band population as a function of center sweep frequency (blue diamonds). A Gaussian fit (blue line) indicates the resonance, which corresponds to the gap between the $S$ and $D$ bands. (\textbf{c}) The experimentally measured band gap is compared to numerical predictions to extract the lattice depth $V_0$ as a function of lattice power setpoint (rust circles), from which we extract a linear fit (rust line) which determines the horizontal error bars in Fig.~\ref{fig:2}b in the main text.}
    \label{sifig:3}
\end{figure}

To calibrate the depth of our optical lattice, we perform amplitude modulation spectroscopy on the $S\rightarrow D$ band transition at zero quasimomentum. To do this, we adiabatically load the BEC from the optical dipole trap into the optical lattice (Fig.~\ref{sifig:3}a) and then perform amplitude modulation with a $\SI{1}{\kilo\hertz}$ frequency sweep over $2\,\mathrm{ms}$. If the band energy difference between the $S$ and $D$ bands at $q=0$ falls into the sweep range, atoms will be excited into the $D$ band; after band mapping, the higher-momentum $D$ band atoms are separated from the $S$ band (zero-momentum) atoms to count their relative populations. Scanning the center frequency of this sweep (Fig.~\ref{sifig:3}b) produces a resonance peak, from which we can extract the band energy difference by fitting. From the fitted band gap, we theoretically calculate the corresponding lattice depth. Linear fitting of the predicted lattice depth against the lattice power PID setpoint results in a lattice depth calibration with respect to the setpoint (Fig.~\ref{sifig:3}c).

\subsection{Force}

\begin{figure}
    \centering
    \includegraphics[width=\linewidth]{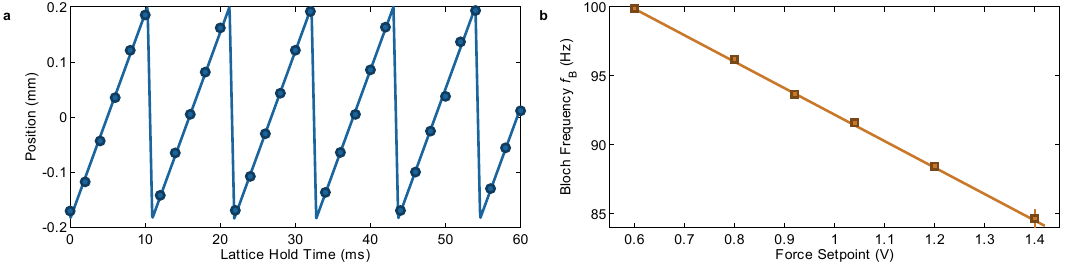}
    \caption{\textbf{Force Calibration.} (\textbf{a}) Fitted atomic positions (blue circles) after band-mapping during a Bloch oscillation with frequency $91.6\pm0.2\,\mathrm{Hz}$ in a static lattice with variable hold time. The triangle-wave fitting curve (blue line) extracts the Bloch frequency. (\textbf{b}) Fitted Bloch oscillation frequency as a function of gradient coil setpoint (rust squares). Vertical error bars result from the aforementioned triangle-wave fits. The linear fit (rust line) determines the horizontal error bars in all the interferometric fringes with respect to Bloch period.}
    \label{fig:Bloch_Frequency_Measurement}
\end{figure}

To measure the force effecting a Bloch oscillation, it suffices to measure the Bloch oscillation frequency $f_\mathrm{B}=1/T_\mathrm{B}=\mathcal{F}/2\hbar k_\mathrm{L}$. To do this, we prepare the atoms in the ground band of a deep ($12.3\,E_\mathrm{R}$) static lattice, initiate a Bloch oscillation with the gradient coils at the chosen setpoint, and measure the quasimomentum using band-mapping after a variable hold time in the lattice. The Bragg scattering at the Brillouin zone edge combined with the nearly-constant dispersion relation of the $S$ band in the deep lattice (tight-binding) limit gives rise to a  sawtooth-like dependence of momentum on the lattice hold time (Fig.~\ref{fig:Bloch_Frequency_Measurement}a), from which we can extract the Bloch oscillation frequency. Performing this same measurement across a range of force setpoints results in a linear fit (Fig.~\ref{fig:Bloch_Frequency_Measurement}b) with which to determine the Bloch frequency and thus the real force over the range of setpoints used for the interferometry experiments. 

\subsection{Modulation Depth}

To calibrate the beamsplitting fraction of the two Landau-Zener avoided crossings, we ramp down our $P$-$D$ modulation once the atoms reach the Floquet-Brillouin zone edge at $q=\hbar k_\mathrm{L}$ so that they undergo only the first avoided crossing. We perform the remainder of the normal interferometer experimental sequence and measure the two output port populations. We empirically select the modulation depth for which the output populations are equal, ensuring that both beamsplitters in the full interferometer sequence set the Landau-Zener transition probability $p=\exp\left(-2\pi\delta\right)=0.5$.

However, we only calibrate this modulation depth for the magic condition; since we fix $\delta V$, the beamsplitting fraction $P$ changes slightly for other lattice depths as a result of changes to $|\varphi_{n,q_\mathrm{r}}\rangle$ (see Eq. \ref{eqn:band_gap}). This undoubtedly accounts for some of the loss in fringe contrast away from the magic condition (Fig.~\ref{fig:2}b in the main text).

\begin{figure}
    \centering
    \includegraphics[width=\linewidth]{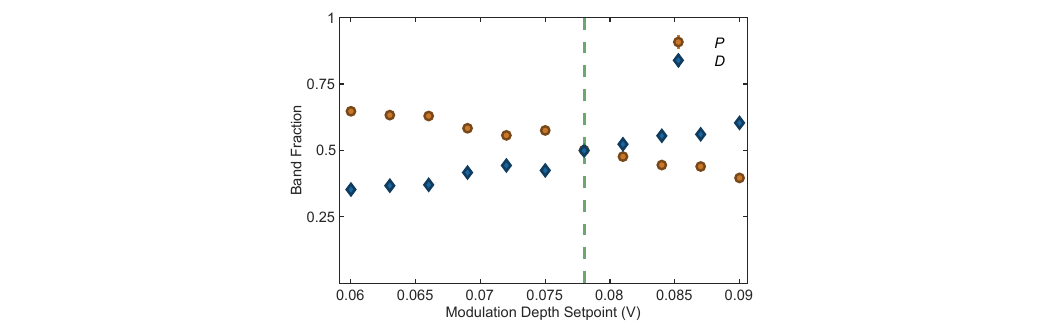}
    \caption{
    \textbf{Calibration of modulation depth}. Relative $P$ and $D$ band populations after the atoms undergo Landau-Zener tunneling at the first avoided crossing. By scanning the modulation depth, we identify the point at which the population split between the two bands is equal. The green dashed line indicates the calibrated modulation depth $\delta V = 0.35\,E_\mathrm{R}$, where our Landau-Zener beamsplitter is effectively set to a 50-50 splitting. Measurements taken at a lattice depth of $V_0 = 8.45\,E_\mathrm{R}$ and modulation frequency $127.4383\,\mathrm{kHz}$. The modulation depth set point is scanned over a range which corresponds to $\delta V = 0.28\,E_\mathrm{R}$ to $\delta V = 0.40\,E_\mathrm{R}$.}
    \label{fig:modDepth}
\end{figure}

\section{Stabilization}

Key parameters of the experiment are actively stabilized during interferometer operation; here we present some details of those feedback loops.

\textbf{Lattice Laser Power:} Our optical lattice beam is produced by an acousto-optic modulator (AOM), so its optical power is controlled by the amplitude of the AOM's RF drive. Since our interferometric phase is sensitive to the lattice depth, we stabilize the laser power using PID feedback from a large-area photodiode. We filter the photodiode output through a $10\,\mathrm{kHz}$ low-pass, since the PID is active during our $\sim100\,\mathrm{kHz}$ amplitude modulation.

\begin{figure}
    \centering
    \includegraphics[width=9cm]{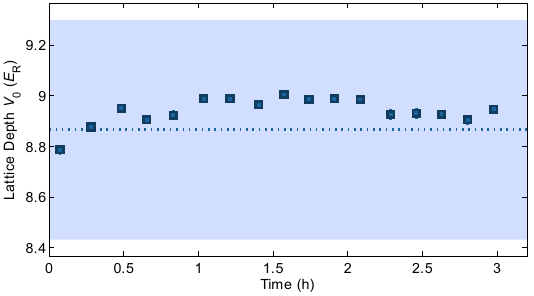}
    \caption{\textbf{Lattice Depth Stability.} Lattice depth as measured by the $S$-$D$ modulation spectroscopy calibration procedure over three hours (blue squares), compared to the numerically predicted magic depth (blue dotted line) and its $\pm\pi/4$ interferometer phase tolerance (light blue region) for a $\Delta q = 0.4 \hbar k_\mathrm{L}$ loop. Error bars arise from the fit of the spectroscopy resonance.}
    \label{sifig:depth_Stability}
\end{figure}

\textbf{Lattice Laser Pointing:} While we do not use a cavity to stabilize our optical lattice mode, we do perform active drift correction of optical lattice pointing. This is done with an active beam stabilization system from MRC Systems GmbH for the incoming lattice beam and its retroreflection; each beam's position is measured by a position-sensitive device (PSD) sensor and steered by a piezo-actuated mirror mount. Between experimental runs, the optical lattice is flashed on for $250\,\mathrm{ms}$ for the MRC PID to recalibrate the pointing of the input and retroreflected beams of the optical lattice, in turn. Careful alignment into the sensors during setup avoids the need to re-align the beams manually.
As shown in Fig.~\ref{sifig:depth_Stability}, the combination of lattice laser power and pointing stabilization gives rise to a stable lattice depth with a drift less than $0.1\, E_\mathrm{R}$ after the initial warm-up stage. This data set was measured via a series of $S$-$D$ modulation spectroscopy scans (as demonstrated in Fig.~\ref{sifig:3}) over three hours.

\textbf{Gradient Coil Current:} The force responsible for Bloch oscillations is provided by a magnetic field gradient produced by a coil whose current is actively stabilized. The current is measured by a transducer (Danisense DS50UB-10V) with a feedback loop implemented through a PI controller (Newport LB1005-S). The controller stabilizes the coil's current using a shunt MOSFET (IXYS IXFN140N20P) in parallel with the coil.

\section{Fringe Contrast Reduction}
A current limitation of interferometer performance is that the fringe contrast decreases as the enclosed space-time area grows. Numerical simulations of a $\Delta q=\hbar k_{\rm L}$ loop that incorporate the axial Rayleigh range of the lattice laser beam and mean-field interactions show no significant change in fringe visibility, indicating that these effects are not responsible for the observed low contrast. For the largest loop ($\Delta q=\hbar k_{\rm L}$), the output imbalance fluctuates around $\mathcal{I} = 0\pm0.1$ instead of showing shot-to-shot noise between $\mathcal{I} = \pm 1$. Therefore, we likewise exclude noise in the lattice depth or the applied force at frequencies below the inverse loop period as a dominant mechanism.

We conclude that the most likely factor leading to reduced contrast is imperfect loop closure. A non-uniform applied force will differentially accelerate atoms in the $P$ and $D$ bands, so that the two arms of the interferometer no longer intersect perfectly at the second beamsplitter. The resulting spatial mismatch is negligible for small loops but increases with enclosed space-time area, naturally explaining the observed contrast decay at large $\Delta q$. Measurements of a $\SI{2}{\hertz}$ residual magnetic field curvature along the lattice direction $x$ and accompanying numerical simulations confirm that the axial field curvature has a negligible effect, leaving transverse inhomogeneities as the leading candidate. Time-of-flight measurements suggest that transverse forces are on the order of $Mg$ along the $y$ and $z$ axes. Such a large magnetic force is likely non-uniform across the millimeter extent that the atoms travel, and even a small gradient may be sufficient to drive transverse excitations leading to a differential position shift between condensates at the end of the interferometer loop, resulting in imperfect closure. If this is the case, identifying and compensating for these transverse gradients should improve contrast for larger loop areas and pave the way for further scaling of the interferometer and its performance. 


\end{document}